\documentclass[aps,pra,twocolumn,showpacs,superscriptaddress]{revtex4-1}

\usepackage{amsfonts,mathrsfs,amsmath,amsthm,amssymb,bbold}                   
\usepackage{epsfig}
\usepackage{bm}
\usepackage{xfrac}
\usepackage{tabulary}
\usepackage{color}
\usepackage[dvipsnames]{xcolor}
\usepackage{graphicx}
\usepackage{slashbox}
\usepackage{hf-tikz}

\begin{document}  
\title {\bf Multiparameter quantum estimation under dephasing noise}

\author{ Le Bin Ho} 
\thanks{Electronic address: binho@kindai.ac.jp\\
Current address: Research Institute of Electrical Communication, Tohoku University, Sendai 980-8577, Japan}
\affiliation{Department of Physics, Kindai University, 
Higashi-Osaka, 577-8502, Japan}

\author{Hideaki Hakoshima}
\affiliation{Nanoelectronics Research Institute, 
National Institute of Advanced Industrial Science and Technology (AIST), 
Ibaraki 305-8568, Japan}

\author{Yuichiro Matsuzaki}
\affiliation{Nanoelectronics Research Institute, 
National Institute of Advanced Industrial Science and Technology (AIST), 
Ibaraki 305-8568, Japan}

\author{Masayuki Matsuzaki}
\affiliation{Department of Physics, Fukuoka University of Education, 
Munakata, Fukuoka 811-4192, Japan}

\author{Yasushi Kondo}
\affiliation{Department of Physics, Kindai University, 
Higashi-Osaka, 577-8502, Japan}
\affiliation{Interdisciplinary Graduate School of Science and Engineering, 
Kindai University, Higashi-Osaka, 577-8502, Japan}

\date{\today}

\begin{abstract}
Simultaneous quantum estimation of multiple parameters 
has recently become essential in quantum metrology. 
Although the ultimate sensitivity of a 
multiparameter quantum estimation 
in noiseless environments can beat 
the standard quantum limit that 
every classical sensor is bounded by,
it is unclear whether the quantum sensor 
has an advantage over the classical one under realistic noise.
In this work, we present a framework of the simultaneous 
estimation of multiple parameters with 
quantum sensors in a certain noisy environment. 
Our multiple parameters to be estimated are 
three components of an external 
magnetic field, and  we consider the noise that causes only dephasing. 
We show that there is an optimal sensing time in the noisy 
environment and the sensitivity can beat the standard quantum limit 
when the noisy environment is non-Markovian. 
\end{abstract}

%
%
\maketitle

\section {Introduction}\label{seci}

Quantum estimation theory is a mathematical framework 
behind quantum metrology and is important for scientific researches 
and technological applications.
Some of its demanded tasks are
minimizing the uncertainty of the estimation 
and attaining an ultimate bound imposed by 
fundamental laws of quantum mechanics.

Great efforts both in theoretical and experimental works 
have been devoted to single-parameter estimation 
\cite{PhysRevLett.102.100401,PhysRevLett.79.3865,PhysRevA.46.R6797,PhysRevA.50.67,
Giovannetti1330,PhysRevLett.96.010401,Jones1166,PhysRevA.82.022330,
Zaiser2016,PhysRevLett.120.140501,PhysRevLett.114.210801,PhysRevLett.113.030401,PhysRevA.92.012120,
PhysRevLett.115.120401,Jordan2015,HO2019153}.  
One of the practical applications of the single parameter estimation 
is to measure external fields such as magnetic fields or electric fields.
When the resonance of a solid-state qubit is shifted by external fields,
we can use a superposition state of the qubit to estimate the amplitude 
of the external fields with a Ramsey type measurements.
With the use of $N$ individual qubits, we can decrease the uncertainty 
of the estimation by $\delta \phi = \mathcal{O} (N^{-\frac{1}{2}})$, 
which is called the standard quantum limit (SQL).
(Here, $\phi$ is a single estimated parameter.)
Moreover, by exploiting entanglement among $N$ qubits, 
we can in principle obtain $\delta \phi =\mathcal{O} (N^{-1})$ 
in the ideal circumstance, and this scaling is called 
the Heisenberg limit (HL) 
\cite{Giovannetti1330,PhysRevLett.96.010401,PhysRevLett.102.100401}.

However, since the entanglement is fragile against decoherence, 
it is not trivial whether the entanglement is useful to decrease 
the uncertainty of the estimation under the effect of realistic noise.
 The effect of noise in the cases of single-parameter estimations 
 has been theoretically \cite{Zhou2018,PhysRevLett.79.3865,
 PhysRevA.84.012103,PhysRevLett.120.140501,PhysRevLett.112.080801,
 PhysRevLett.115.200501,PhysRevLett.112.150801,PhysRevLett.112.150802,
 matsuzaki2017magnetic,PhysRevX.7.041009,Zhou2018}
 and experimentally \cite{Taylor2008,Jones1166,PhysRevLett.106.080802,
 PhysRevA.94.012324,PhysRevLett.116.230502,Ho_2019,ho2019nuclear}
 discussed. 
The most important noise for the solid-state qubits is the dephasing one. 
It is known that one cannot beat 
the SQL for the estimation of the field amplitude
under the effect of Markovian dephasing noise
even with the use of the entanglement \cite{PhysRevLett.79.3865}. On the other hand, 
recent studies show that,  if the dephasing noise has 
non-Markovian properties, 
one can obtain the scaling of $\delta \phi = \mathcal{O} (N^{-3/4})$ 
by using the entanglement for single parameter estimation, 
and this scaling beats the SQL 
\cite{Jones1166, PhysRevA.84.012103,chin2012quantum,macieszczak2015zeno,
tanaka2015proposed,smirne2016ultimate,haase2018fundamental}. 
The crucial feature of the non-Markovian noise is to show a quadratic behavior 
as a function of time at the initial decay, 
which is called a Zeno regime, and the interaction time
between the entanglement and target fields is adjusted in this regime 
to obtain the quantum enhancement of the sensitivity
\cite{Jones1166, PhysRevA.84.012103,chin2012quantum,macieszczak2015zeno,
tanaka2015proposed,smirne2016ultimate,haase2018fundamental,dooley2016hybrid}. 
For the estimation of the amplitude of the field, 
$\delta \phi =\mathcal{O} (N^{-3/4})$ 
is considered as the ultimate scaling 
under the effect of the non-Markovian dephasing noise 
\cite{macieszczak2015zeno,smirne2016ultimate,
haase2018fundamental}.

On the other hand, great attention has been paid to
multiparameter estimations 
\cite{PhysRevLett.123.200503,PhysRevLett.116.030801}. 
For example, estimations of phase and phase diffusion (loss) 
\cite{Vidrighin2014,PhysRevA.92.032114,Szczykulska_2017,
knysh2013estimation,PhysRevA.89.023845,PhysRevA.88.040102,
PhysRevA.96.062306,Roccia:18},
phase-space displacements \cite{PhysRevA.87.012107,Steinlechner2013}, 
multiple phases 
\cite{Cyril2013,PhysRevLett.111.070403,Liu_2017,PhysRevLett.116.030801},
damping and temperature \cite{PhysRevA.83.012315}, 
waveforms \cite{PhysRevX.5.031018}, and
operators \cite{PhysRevA.65.012316,PhysRevA.69.022303}.
One of the practical applications of the multiparameter estimation 
is to measure vector magnetic fields.
Imaging of the vector magnetic fields from the biomaterials or  
circuit current 
is especially important for the medical and materials science, 
and these have been discussed and demonstrated 
\cite{le2013optical,nowodzinski2015nitrogen}.

In this work, we numerically investigate 
the multiparameter estimation under the influence of dephasing noise. 
In particular, we consider the case to estimate 
three vector components of the target fields
by using the entanglement
under the effect of dephasing noise.  
Moreover, we study the performance of the entangled sensor for 
multiparameter estimation under both Markovian and non-Markovian 
dephasing noises. Although numerical calculations 
of noisy quantum systems with many qubits
are difficult because the size of the density matrix 
grows exponentially as the number of the qubits increases,
the recent studies show that 
the cost for the calculation is tractable when the qubits 
are identical two-level systems
\cite{shankar2017steady,kirton2017suppressing,PhysRevA.98.063815}. 
We adopt this technique, and numerically 
calculate the uncertainty of the estimation 
to check how the uncertainty scales as a function of the number of the qubits.
We show that, under the effect of non-Markovian dephasing noise, 
we can beat the SQL for the multiparameter 
estimation, and the scaling that we obtain by fitting 
the numerical results 
$\delta \bm\phi =\mathcal{O} (N^{-3/4})$, 
which is the same as that of the ultimate scaling 
for the single parameter estimation
under dephasing noise.
(Here, vector $\bm\phi$ is a set of multiple parameters.) 
Our analysis would provide further understanding 
of quantum metrology.

This paper is organized as follows. 
Section~\ref{secii} introduces our measurement framework
estimating multiple parameters simultaneously. 
The numerical results are presented in Sec.~\ref{seciv}. 
We summarize our work in Sec.~\ref{secv}.

\section{Multiparameter estimation framework}\label{secii}
\subsection{Dynamics of an $N$-identical particles sensor}

We consider a sensor consisting of an ensemble of 
$N$-identical two-level systems.
The two-level system at the $n$th site  
can be characterized by the Pauli operators as 
$J_{\alpha}^{(n)} = \frac{1}{2}\sigma_{\alpha}^{(n)}$ for
$\alpha = \{x, y, z\}.$ 
The whole sensor operators are given as
$J_\alpha = \sum_nJ_{\alpha}^{(n)}$.
To be concrete, we assume that  these two-level systems are   
one-half spins and that the field to be sensed is a magnetic field.  

The sensor dynamics without noise is governed by the Hamiltonian 
\begin{align}
\label{eq:H}
H(\bm\phi) = \phi_x J_x+\phi_y J_y+\phi_z J_z,
\end{align}
where a set of three parameters
$\bm{\phi} = (\phi_x,\phi_y, \phi_z)$
describes the magnetic field to be estimated. 
Also, this magnetic field provides the quantization axis of each qubit.
We assume the sensor is governed by the GKLS 
master equation \cite{doi:10.1142/S1230161217400017}, 
\begin{align}\label{eq:fns}
\frac{d \rho_t(\bm\phi)}{dt} 
= -i [H(\bm\phi), \rho_t(\bm\phi)]+ {\mathcal L}[\rho_t(\bm\phi)],
\end{align}
where $\rho_t(\bm\phi)$ is the quantum state of the sensor at time $t$,
and we take the natural unit system, or $\hbar =1$. 
Further, we assume the followings
\begin{align}
\label{eq:lin}
{\mathcal L}[\rho_t(\bm\phi)] = -\gamma_t
\sum_{n=1}^N 
\bigl[a^{(n)},[a^{(n)},\rho_t(\bm\phi)]\bigr],
\end{align}
where $\gamma_t$ characterizes the strength of 
the noise and 
\begin{align}
\label{eq:an}
a^{(n)} = \bm{\varphi \cdot J}^{(n)} = 
\varphi_xJ_x^{(n)} +  \varphi_yJ_y^{(n)} +  \varphi_zJ_z^{(n)},
\end{align}
where $a^{(n)}$ is the operator acting on the $n$th-site spin and
is normalized so that 
$[a^{(n)}]^2=\bm{I}$, or 
$\varphi_x^2+\varphi_y^2+\varphi_z^2 = 4$. 

Then, we consider dephasing noise by assuming $\bm\phi \parallel \bm\varphi$
where the environmental noisy fields are applied along the quantization axis
of the system. A similar noise has been considered 
in single parameter estimation in 
Refs.~\cite{PhysRevLett.123.200503,PhysRevA.84.012103,chin2012quantum}. 
This assumption leads to the property such that $H(\bm\phi)$ and $a^{(n)}$ 
commute and thus the sensor dynamics calculation becomes tractable.
Such a dephasing noise is often considered as a dominant noise 
in the solid-state systems and NMR. 

A Markovian and 
non-Markovian noisy environment can be introduced by 
taking the noise strength $\gamma_t$ as 
\begin{align}
\gamma_t = \left\{
  \begin{array}{ll}
    \gamma & : \text{Markovian}\\
    \gamma^2 t & : \text{non-Markovian}
  \end{array}
\right. .
\end{align}

We provide a detailed calculation for 
the dynamics of such a sensor in Appendices~\ref{appA} and \ref{appB}.

\subsection{The precision of the estimation}
The precision of the estimation of $\bm\phi$ 
is evaluated by its covariance matrix, 
$[\bm{V} (\bm{\phi})]_{\alpha,\beta} = 
\langle \phi_\alpha \phi_\beta \rangle-
\langle \phi_\alpha \rangle \langle \phi_\beta\rangle$.
The diagonal elements $[\bm{V}(\bm{\phi})]_{\alpha,\alpha}$ are 
the variance $(\delta\phi_\alpha)^2$
while the off-diagonal elements are the correlations 
between different parameters. 
The quantum Cram\'er-Rao bound 
is a lower bound to the covariance matrix in terms of 
the  classical Fisher information matrix (CFIM)
and quantum Fisher information matrix (QFIM),  such that
\begin{align}\label{eq:lb}
M\cdot\bm{V}(\bm{\phi}) \ge [{\bm F}(\bm\phi)]^{-1}
\ge [{\bm Q}(\bm\phi)]^{-1},
\end{align}
where $M$ is the number of repeated measurements in the 
total measurement time $T$,
$\bm{F}$ and $\bm{Q}$ are the CFIM and QFIM, respectively. 
The first inequality is a classical Cram\'er-Rao bound
(CCRB), while the second one is referred to as a quantum 
Cram\'er-Rao bound (QCRB).
The CFIM is 
given through the measurement probabilities 
$
[{\bm F}(\bm\phi)]_{\alpha\beta} = 
\sum_l\frac{1}{P(l|\bm{\phi})}
\left[\partial_{\alpha} P(l|\bm{\phi})\right]
\left[\partial_{\beta} P(l|\bm{\phi})\right],
$
where $\{\alpha, \beta\} = \{x, y, z\}$ and 
$P(l|\bm{\phi}) = {\rm Tr}[\Pi_l\rho_t(\bm\phi)]$ 
determined by a POVM $\{\Pi_l\}$ and where we have used 
$\partial_{\alpha}\rho_t(\bm\phi) \equiv
\frac{\partial\rho_t(\bm\phi)}{\partial\phi_\alpha}$ for short.
When $\rho_t(\bm{\phi}) $ is able to be spectral decomposed 
so that $\rho_t(\bm{\phi}) =  \sum_l p_l  | l \rangle \langle l |$, 
the QFIM is given by 
\begin{align}\label{eq:Qsol_rec}
[\bm Q(\bm\phi)]_{\alpha,\beta} = 
2\sum_{p_l+p_{l'}>0}
\dfrac{\langle l|\partial_{\alpha}\rho_t(\bm\phi)|l'\rangle
\langle l'|\partial_{\beta}\rho_t(\bm\phi)|l\rangle}
{p_l + p_{l'}}.
\end{align}
Although the number of $l$ is exponentially large ($2^N$), 
we can reduce the calculation cost 
of which order is $N^2$ when the qubits are symmetric in terms 
of permutation operations
(See Appendix~\ref{appA} for details).

From the trace of Eq.\ (\ref{eq:lb}), 
we will analyze the lower bound of the total variance  
\begin{align}\label{eq:total_v}
|\delta\bm\phi|^2 \ge {\rm Tr}\bigl[[\bm Q(\bm\phi)]^{-1}\bigr]/M,
\end{align}
where $ |\delta\bm\phi|^2 \equiv 
{\rm Tr}\bigl[\bm{V}(\bm{\phi})\bigr]$ 
is the total variance, 
which is the summation of three partial variances,
i.e., $|\delta\bm\phi|^2 = |\delta\phi_x|^2 +
|\delta\phi_y|^2+|\delta\phi_z|^2$.
The lower bound in the R.H.S. of Eq.~\eqref{eq:total_v}
is the ultimate bound that all three components 
can be achieved simultaneously. 

\section{Numerical results}\label{seciv}

\subsection{Simultaneous versus individual scenarios}
\begin{figure} [t!]
\includegraphics[width=8cm]{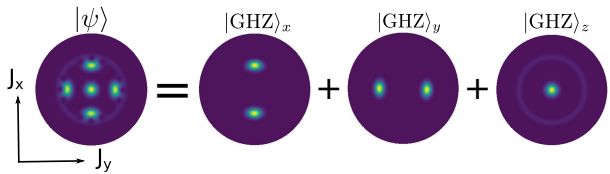}
\caption{
(Color online) The visualization of Husimi functions 
for $|{\rm GHZ}\rangle_k$, 
$(k = x, y, z)$ and $|\psi\rangle$ given in Eq. \eqref{eq:psi}.
We fixed $N = 40$.}
\label{fig:WF_xyz}
\end{figure}

We consider two scenarios for the estimation: 
simultaneous estimation and individual estimation.
For the simultaneous scenario, three components 
of the field 
will be estimated simultaneously. 
The initial state is set to be
 $\rho_{t=0} = |\psi\rangle\langle\psi|,$
where
\begin{align}\label{eq:psi}
|\psi\rangle = \mathcal{N}
\bigl(|{\rm GHZ}\rangle_x + |{\rm GHZ}\rangle_y + |{\rm GHZ}\rangle_z\bigr),
\end{align}
$\mathcal{N}$ is the normalization constant.
The GHZ state is defined as
\begin{align}\label{eq:GHZi}
|{\rm GHZ}\rangle_k = 
\dfrac{|\lambda^{\rm max}_k\rangle+|\lambda_k^{\rm min}\rangle}
{\sqrt{2}},
\end{align}
where $|\lambda^{\rm max}_k\rangle$ and 
$|\lambda^{\rm min}_k\rangle$ are the two eigenstates
of $J_k$ (${k = x, y, z}$) that correspond to the maximum 
and minimum eigenvalues $\lambda^{\rm max}_k$
and $\lambda^{\rm min}_k$, respectively. 
If there is no noise, 
an entanglement sensor using the state $|\psi\rangle$  
provides the Heisenberg scaling
for the multiparameter estimation 
as shown in 
Ref.~\cite{PhysRevLett.116.030801}.
In Fig. \ref{fig:WF_xyz}, we visualize the Husimi function 
\cite{1940264}
of the three GHZ states and 
$|\psi\rangle$ for $N=40$.

For the individual scenario, each component will be estimated 
separately after repeated $M/3$ measurements.
In this case, we use an entangled state $|{\rm GHZ}\rangle_k$ in 
Eq.~\eqref{eq:GHZi} (${k = x, y, z}$) 
to measure the corresponding magnetic field. 
This scheme is a direct application 
of the single parameter estimation to the vector field sensing.

We define the lower bound of the total variance in
Eq.~\eqref{eq:total_v} for the simultaneous scenario as
\begin{align}
\label{eq:I}
\mathcal{I}_{\rm sim} \equiv {\rm Tr}\bigl[[\bm Q(\bm\phi)]^{-1}\bigr]/M,
\end{align}
where `sim' stands for `simultaneous.'
Although the use of such a lower bound may not provide
full insight into the variance, 
we emphasize that such a lower bound can be in principle achieved 
by using an optimal minimization scheme such as SDP 
\cite{PhysRevLett.123.200503} 
or a general JNT-QEC \cite{gorecki2019optimal}.

For the individual scenario, we define the total variance as
\begin{align}
\mathcal{I}_{\rm ind} = \dfrac{3}{M}\Bigl(Q_x^{-1} + 
Q_y^{-1}+Q_z^{-1}\Bigr),
\end{align}
where $Q_k = [\bm Q(\bm\phi)]_{k,k}$ (${k = x, y, z}$) 
setting the initial state $|{\rm GHZ}\rangle_k$.
Here, $\frac{3}{M}$ denotes $M/3$ 
repeated measurements devoted to estimating
a component $k$.

We emphasize that our framework here is
different from Ref.~\cite{PhysRevLett.116.030801}.
While Ref.~\cite{PhysRevLett.116.030801} studies the 
noiseless case,
here we have extended its calculation technique for
a sensor under noise. 

\subsection{The total variance under the dephasing noise}
Here, we investigate the performance of 
the entangled sensor for the multiparameter estimation
under the effect of the dephasing noise.
To examine the numerical results, we fix
$\bm\varphi = (2/\sqrt{3}, 2/\sqrt{3}, 2/\sqrt{3})$
in Eq.~\eqref{eq:an}
and $\bm\phi = (0.01, 0.01, 0.01)$  in Eq.~\eqref{eq:H}.
Here, we assume, as usual, that the necessary time 
for the state preparation and readout is negligibly small. 
We fix the total time $T = 100$ 
and investigate $\mathcal{I}$ 
for Markovian and non-Markovian cases. 

\begin{figure} [t]
\includegraphics[width=8.65cm]{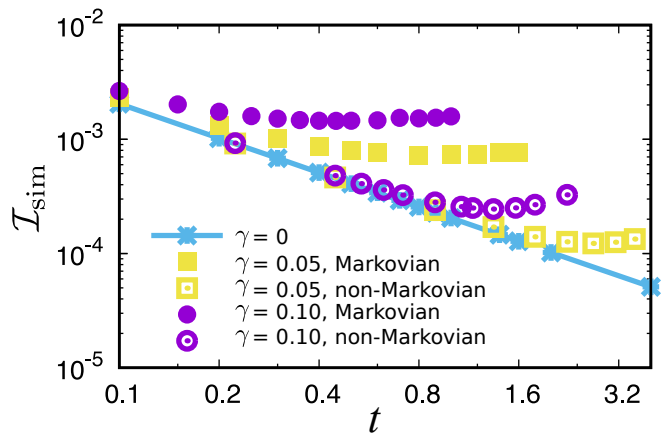}
\caption{(Color online) 
The plot of $\mathcal{I}_{\rm sim}\equiv 
{\rm Tr}\bigl[[\bm Q(\bm\phi)]^{-1}\bigr]/M$
as a function of measurement time $t$ at $T=100$ and 
$N=20$ at  $\gamma = 0.1$ 
and 0.05 
in both Markovian 
and non-Markovian 
dephasing noisy environment. ${\mathcal I}_{\rm sim}$'s 
at $\gamma=0$ are plotted for comparison. 
Note that $M = T/t$. 
}
\label{fig:figQFIM_N20_fs}
\end{figure}

Figure~\ref{fig:figQFIM_N20_fs} shows 
$\mathcal{I}_{\rm sim}$
as a function of measurement time $t$ for $N = 20$.
Note that we are allowed to measure for $T$ and thus 
$M = T/t$ in Eq.~(\ref{eq:I}).  
We investigate $\gamma = 0, 0.05$, and $0.1$ cases. 
In the absence of noise ($\gamma = 0$), 
the longer $t$ always gives the better measurements. 
When the noise is present ($\gamma \ne 0$),
we found that there are minima of $\mathcal{I}_{\rm sim}$
as a function of $t$: There are optimal measurement 
times $t^{\rm opt}$'s as functions of $N$. $t^{\rm opt}$
in the case of Markovian noise is shorter than that in 
the case of non-Markovian one.

\begin{figure} [t]
\includegraphics[width=8.65cm]{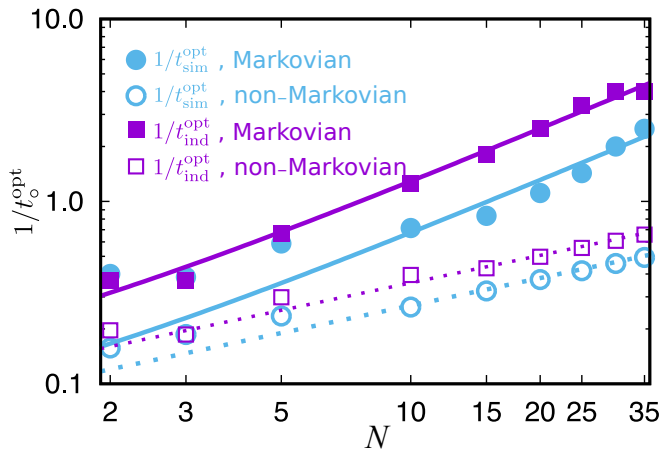}
\caption{(Color online) 
$1/t^{\rm opt}_\circ$ as a function of $N$ 
for two cases of Markovian and non-Markovian 
dephasing noisy environments. 
Here $\circ$ = sim or ind.
The dotted lines 
show $\sqrt{N}$ dependence, while the solid
lines do $N$ dependence. We fit the data for $N\ge10$. 
}
\label{fig:t_opt}
\end{figure}

\begin{figure} [t]
\includegraphics[width=8.6cm]{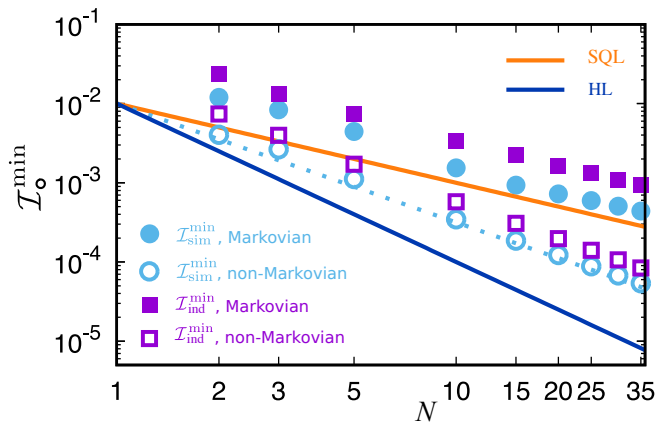}
\caption{(Color online) 
$\mathcal{I}_\circ^{\rm min}$ 
as a function of $N$ in the Markovian and non-Markovian 
dephasing noisy environments 
when $\gamma = 0.05$. To show the SQL and HL 
dependences, $1/TN$ (orange line, SQL) 
and  $ 1/TN^{2}$ (blue line, HL) are plotted. 
The line $1/ TN^{1.5}$ 
is also plotted (cyan dotted line).
}
\label{fig:figQFIM_fN}
\end{figure}

We investigate $t^{\rm opt}_{\circ}$ as a function of $N$  for both 
simultaneous ($\circ$ = sim) and individual ($\circ$ = ind) scenarios 
in both Markovian and non-Markovian cases. 
Figure~\ref{fig:t_opt} shows that $1/t^{\rm opt}_{\circ}$ is proportional to $N$ ($\sqrt{N}$) 
in the Markovian (non-Markovian) case at $N \ge 10$, as expected
\cite{chin2012quantum,PhysRevLett.79.3865}. We found, however, that $t^{\rm opt}_{\rm sim}$'s behave 
differently at $N<10$ in both Markovian and non-Markovian cases. We suspect 
that $N<10$ is too small to observe the expected dependences. 
These observations are consistent with $N$ dependences of 
${\mathcal I}^{\rm min}_{\circ}$ in Fig.~\ref{fig:figQFIM_fN}. 

Figure~\ref{fig:figQFIM_fN} shows $\mathcal{I}_{\rm \circ}^{\rm min}$
for $\gamma = 0.05$
at $t = t^{\rm opt}_\circ$, as a function of $N$.  
Here $\circ$ = sim or ind.  
We observe the followings for the Markovian case. 
(i) $\mathcal{I}_{\rm \circ}^{\rm min}$ becomes proportional to 
$N^{-1}$ at $N \ge 10$ or has the same dependence with the SQL 
and thus the entangled sensor has no benefit, and 
(ii) $\mathcal{I}_{\rm ind}^{\rm min} 
> \mathcal{I}_{\rm sim}^{\rm min}$ at the same $N$ 
which implies that the simultaneous measurement is 
beneficial. 
Those behaviors are consistent with the case of 
the single parameter estimation where
entangled sensors cannot beat the SQL
 under the effect of the Markovian dephasing noise 
 \cite{PhysRevLett.79.3865}.

In contrast, we observe the followings for the non-Markovian case. 
(i) $\mathcal{I}_\circ^{\rm min}$ is proportional to $N^{-1.5}$ at $N\ge 10$
and thus the entangled sensor is beneficial, and 
(ii) $\mathcal{I}_{\rm ind}^{\rm min} 
> \mathcal{I}_{\rm sim}^{\rm min}$ (although this difference is small)
at the same $N$ which implies that the simultaneous measurement is 
beneficial. The observation (i) ($N^{-1.5}$ dependence) 
is a well-known scaling for non-Markovian dephasing
\cite{PhysRevA.84.012103,chin2012quantum}.
The observation (ii) was reported in Ref. \cite{PhysRevLett.116.030801} 
for a noiseless case. Whereas, in this work, 
we show the reduction of uncertainty in noisy cases.

\section{Conclusion}\label{secv}
In conclusion, we analyze the simultaneous estimation 
of the multiple parameters 
with an entangled sensor in both Markovian and non-Markovian 
dephasing noisy environment. We found that the entangled 
sensor is beneficial in the non-Markovian environment 
while it is not the case in the Markovian one.

Our multiple parameters are the components of a magnetic field 
and are sensed with an ensemble of $N$-identical particles 
that are entangled with each other. 
By taking into account the symmetry in permutation operators, 
the calculation cost is drastically reduced and becomes tractable.
The entangled sensor is exposed 
to the target fields under the effect of the dephasing noise. 
We numerically calculate the quantum Fisher information matrix 
and investigate the lower bound of the total variance, 
denoted as $\mathcal{I}$. 
When the dephasing noise is present, it always prevents us 
from achieving the Heisenberg limit. We, however, found that 
an entangled sensor can beat the standard quantum limit in a non-Markovian 
dephasing noise but not in a Markovian noise.


\begin{acknowledgments}
This work was supported by CREST(JPMJCR1774), JST.
This work was also supported by Leading Initiative 
for Excellent Young Researchers MEXT Japan, 
MEXT KAKENHI (Grant No. 15H05870), 
and JST presto (Grant No. JPMJPR1919) Japan.
LBH is grateful to Nathan Shammah for useful discussions on 
QuTiP. 
\end{acknowledgments}

\begin{widetext}

\appendix
\setcounter{equation}{0}
\renewcommand{\theequation}{A.\arabic{equation}}
\section{Permutation symmetric sensor}\label{appA}
We consider the sensor
consists of $N$-identical particles where 
the permutation symmetry is taken as follows 
\cite{PhysRevA.98.063815,PhysRevA.78.052101,PhysRevA.81.032104}. 
The joint Hilbert space of the sensor 
is $\mathscr{H}_N = \mathscr{H}^{(1)}
\otimes\cdots\otimes\mathscr{H}^{(N)}$
with ${\rm dim}(\mathscr{H}_N) = 2^N$.
Any quantum state of the sensor can be given as
\begin{align}\label{eq:psi:ensemble}
|\psi\rangle=\sum_{m_1,m_2,\cdots,m_N}
c_{m_1,m_2,\cdots,m_N}
|m_1,m_2,\cdots,m_N\rangle,
\end{align}
where the product basis
$|m_1,m_2,\cdots,m_N\rangle=
|m_1\rangle\otimes|m_2\rangle\otimes\cdots\otimes
|m_N\rangle$,
with $m_n=\pm\frac{1}{2}$
are eigenvalues of $J_z^{(n)}$.
This basis 
is the eigenstate of the spin operators 
$\bm{J}^{(n)}$ and $J_z^{(n)}$
\begin{align}
[\bm{J}^{(n)}]^2|m_1,m_2,\cdots,m_N\rangle
&= j_n(j_n+1)|m_1,m_2,\cdots,m_N\rangle, \label{eq:j2:full}\\ 
J_z^{(n)}|m_1,m_2,\cdots,m_N\rangle
&= m_n|m_1,m_2,\cdots,m_N\rangle. \label{eq:jz:full}
\end{align}

The above product basis can be represented by an irrep basis, 
which consists of the total spin eigenstates
\cite{PhysRevA.78.052101,PhysRevA.81.032104}
\begin{align}
\bm{J}^2|j,m,i\rangle
&= j(j+1)|j,m,i\rangle, \label{eq:j2:irreps}\\ 
J_z|j,m,i\rangle
&= m|j,m,i\rangle, \label{eq:jz:irreps}
\end{align}
where $|j,m,i\rangle$ is the irrep basis,
$j\le N/2$ the total angular momentum, $|m|\le j$. 
For each $j$, the quantum number $i = 1, \cdots d_N^j$,
where 
\begin{align}\label{eq:dNj}
d_N^j = \dfrac{N!(2j+1)}{(N/2-j)!(N/2+j+1)!}
\end{align}
is the number of degenerate irreps for each $j$ \cite{Mihailov_1977}
(the number of ways to combine $N$ particles 
that gets the total angular momentum $j$.)
The coefficient 
$c_{m_1,m_2,\cdots,m_N}$ now becomes
$c_{j,m,i}$.
Taking into account the permutation symmetry where
all the degenerate irreps of each $j$
are indistinguishable, i.e., 
$c_{j,m,i} = c_{j,m,i'} \ \forall i,i' \in [1, d_N^j]$,
then, the irrep basis $|j,m,i\rangle$ can be gathered to
the Dicke basis $|j,m\rangle$ \cite{PhysRev.93.99}, 
where
\begin{align}\label{eq:dicke_basis}
|j,m\rangle = \dfrac{1}{\sqrt{d_N^j}}
\sum_{i=1}^{d_N^j}|j,m,i\rangle.
\end{align}
This basis is the eigenstate of the 
collective pseudo-spin operators
\begin{align}
\bm{J}^2|j,m\rangle &= j(j+1)|j,m\rangle, \label{eq:j2}\\ 
J_z|j,m\rangle &= m|j,m\rangle. \label{eq:jz}
\end{align}
Under this symmetry, the dimension now reduces to 
the Dicke-basis dimension $d_D$:
\begin{align}
d_D = 
  \begin{cases} 
   (N+3)(N+1)/4 & \text{for odd } N, \\
   (N+2)^2/4       & \text{for even } N.
  \end{cases}
\end{align}  
Hereafter, we take $\bm{J}, J_\alpha$ as the
collective pseudo-spin operators in the $d_D$ dimension.

In the  $d_D$ dimension, $J_\alpha$ has a structure 
of block matrices as shown in Fig. \ref{dick_structure}.
The first block corresponds to $j = N/2$, 
the explicit form of this block is a spin-$j$ operator $S_\alpha,
\alpha = \{x, y, z\}$.
The construction for others is the same.
For example, $N = 3$, we have
\begin{align}\label{eq:jx}
J_x = 
\begin{pmatrix}
0 & \sqrt{3}/2  & 0 & 0 & 0 & 0\\
\sqrt{3}/2  & 0 & 2 & 0 & 0 & 0\\
0 & 2 & 0 & \sqrt{3}/2 & 0 & 0\\
0 & 0 & \sqrt{3}/2 & 0 & 0 & 0\\
0 & 0 & 0 & 0 &  0 & 1/2\\
0 & 0 & 0 & 0 & 1/2 & 0
\end{pmatrix}.
\end{align}
Do the same for $J_y$ and $J_z$.
\begin{figure} [t!]
\includegraphics[width=8cm]{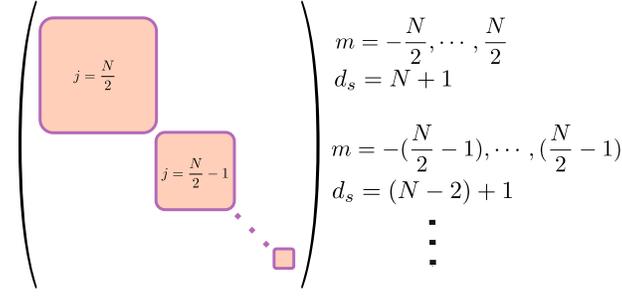}
\caption{
(Color online) Block-diagonal form of a quantum state 
representing in the Dicke basis.
The first block corresponds to $j = N/2$
and its sub-dimension is $d_s = N +1$.
Do the same calculation for 
the remaining blocks in the diagonal matrix.
The off-diagonal terms are all zeros. 
}
\label{dick_structure}
\end{figure} 

\setcounter{equation}{0}
\renewcommand{\theequation}{B.\arabic{equation}}
\section{Dynamic of permutation symmetric sensor under dephasing noise}
\label{appB}
We will solve the GKLS equation 
\eqref{eq:fns} in the main text in $d_D$ dimension. 
We note that $[H(\bm\phi),a^{(n)}] = 0$, 
thus, we first calculate only the Liouville term \eqref{eq:lin}.
The following expressions are independent 
of the choice of the direction of the quantization axis, which 
is physically determined by the target field to be measured. 
We rewrite it here:
\begin{align}\label{eq:linex_appB}
\dfrac{\partial\rho_t}{\partial t} = 2\gamma_t
\Bigl(\sum_{n=1}^N 
a^{(n)}\rho_t a^{(n)}-N\rho_t\Bigr).
\end{align}

We first show how to calculate the Liouvillian superoperator
in the R.H.S. of Eq. \eqref{eq:linex_appB}.
Using $a^{(n)} = \varphi_xJ_x^{(n)} +  \varphi_yJ_y^{(n)} +  \varphi_zJ_z^{(n)}$,
the summation term in Eq. \eqref{eq:linex_appB} is: 
(for short, we first keep $\rho_t$)
\begin{align}\label{eq:sum}
\notag\sum_{n=1}^Na^{(n)}\rho_t a^{(n)} &= \sum_{n=1}^N
\Bigl[\varphi_xJ_x^{(n)} +  \varphi_yJ_y^{(n)} +  \varphi_zJ_z^{(n)}\Bigr]
\rho_t\Bigl[\varphi_xJ_x^{(n)} +  \varphi_yJ_y^{(n)} +  \varphi_zJ_z^{(n)}\Bigr]\\
\notag&=\sum_{n=1}^N
\Bigl[\dfrac{\varphi_x}{2}\bigl(J_+^{(n)}+J_-^{(n)}\bigr) +  
\dfrac{i\varphi_y}{2}\bigl(J_-^{(n)}-J_+^{(n)}\bigr) +  \varphi_zJ_z^{(n)}\Bigr]
\rho_t\Bigl[\cdots\Bigr]\\
&=\sum_{n=1}^N
\Bigl[\varphi_w^*J_+^{(n)} +  \varphi_wJ_-^{(n)}+  \varphi_zJ_z^{(n)}\Bigr]
\rho_t\Bigl[\varphi_w^*J_+^{(n)} +  \varphi_wJ_-^{(n)}+  \varphi_zJ_z^{(n)}\Bigr],
\end{align}
where $J_\pm^{(n)} = J_x^{(n)}\pm iJ_y^{(n)}, 
\varphi_w = (\varphi_x + i\varphi_y)/2.$
Finally, we have
\begin{align}\label{eq:suma}
\notag\sum_{n=1}^Na^{(n)}\rho_t a^{(n)} &= 
\sum_{n=1}^N
\Bigl[
(\varphi_w^*)^2J_+^{(n)}\rho_t J_+^{(n)}
+|\varphi_w|^2J_+^{(n)}\rho_t J_-^{(n)}
+\varphi_w^*\varphi_zJ_+^{(n)}\rho_t J_z^{(n)}\\
\notag&\hspace{1cm}+
|\varphi_w|^2J_-^{(n)}\rho_t J_+^{(n)}
+(\varphi_w)^2J_-^{(n)}\rho_t J_-^{(n)}
+\varphi_w\varphi_zJ_-^{(n)}\rho_t J_z^{(n)}\\
&\hspace{1cm}+
\varphi_w^*\varphi_zJ_z^{(n)}\rho_t J_+^{(n)}
+\varphi_w\varphi_zJ_z^{(n)}\rho_t J_-^{(n)}
+\varphi_z^2J_z^{(n)}\rho_t J_z^{(n)}
\Bigr].
\end{align}
Here, these terms corresponding to 
$J_+^{(n)}\rho_t J_-^{(n)}$,
$J_-^{(n)}\rho_t J_+^{(n)}$, and 
$J_z^{(n)}\rho_t J_z^{(n)}$ are local pumping, local emission,
and local dephasing, respectively. 
Now, using $\rho_t = \sum_{jmm'} p_{jmm'}|j,m\rangle\langle j,m'|.$ 
Then for each $j,m,m'$, we have
\cite{PhysRevA.78.052101,PhysRevA.81.032104,PhysRevA.98.063815}

\begin{align}\label{eq:sum1}
\notag \sum_{n=1}^NJ_{k}^{(n)}|j,m\rangle\langle j,m'| J_{l}^{(n)\dagger} &= 
\notag a_{kl}^N |j,m_k\rangle\langle j,m'_l|\\
\notag &+b_{kl}^N |j-1,m_k\rangle\langle j-1,m'_l|\\
&+d_{kl}^N |j+1,m_k\rangle\langle j+1,m'_l|,
\end{align}
where $k,l =\{+,-,z\}, m_+=m+1, m_-=m-1,m_z=m$, and
\begin{align}\label{eq:abc}
\notag a_{kl}^N &= A_k^{j,m}A_l^{j,m'}\dfrac{1}{2j}
\Bigl(1+\dfrac{\alpha_N^{j+1}}{d_N^j}\dfrac{2j+1}{j+1}\Bigr),\\
\notag     &=A_k^{j,m}A_l^{j,m'}\dfrac{N/2+1}{2j(j+1)}\\
              &:=A_k^{j,m}A_l^{j,m'}\Lambda_a,\\
\notag b_{kl}^N &= B_k^{j,m}B_l^{j,m'}\dfrac{\alpha_N^{j}}{2jd_N^j},\\
\notag               &= B_k^{j,m}B_l^{j,m'} \dfrac{N/2+j+1}{2j(2j+1)}\\
                         &:= B_k^{j,m}B_l^{j,m'} \Lambda_b,\\
\notag d_{kl}^N &= D_k^{j,m}D_l^{j,m'}\dfrac{\alpha_N^{j+1}}{2(j+1)d_N^j},\\
\notag               &=D_k^{j,m}D_l^{j,m'}\dfrac{N/2-j}{2(j+1)(2j+1)}\\
			&:=D_k^{j,m}D_l^{j,m'}\Lambda_d,
\end{align}
where 
\begin{align}\label{eq:ABD}
A_\pm^{j,m} &=\sqrt{(j\mp m)(j\pm m+1)}, \ A_z^{j,m}=m,\\
B_\pm^{j,m} &=\pm\sqrt{(j\mp m)(j\mp m-1)}, \ B_z^{j,m}=\sqrt{(j+m)(j-m)},\\
D_\pm^{j,m} &=\mp\sqrt{(j\pm m+1)(j\pm m+2)}, \ D_z^{j,m}=\sqrt{(j+m+1)(j-m+1)},\\
\Lambda_a &= \frac{N/2+1}{2j(j+1)}, 
\Lambda_b = \frac{N/2+j+1}{2j(2j+1)},
\Lambda_d = \frac{N/2-j}{2(j+1)(2j+1)},
\end{align}
and 
\begin{align}\label{eq:al}
\alpha_N^j = \sum_{j'=j}^{N/2}d_N^{j'}=\dfrac{N!}{(N/2-j)!(N/2+j)!},
\end{align}
with the degenerate
$d_N^j = \dfrac{N!(2j+1)}{(N/2-j)!(N/2+j+1)!}$.

We calculate explicitly Eq. \eqref{eq:sum1} for each $j, m, m'$, where
\begin{align*}
\varphi_z^2\sum_{n=1}^N J_z^{(n)}|j,m\rangle\langle j,m'|J_z^{(n)} 
&= \varphi_z^2 \bigl(mm' \Lambda_a 
|j,m\rangle\langle j,m'|  & \leftarrow \Gamma^{(1)}\\
&+ B_z^{j,m}B_z^{j,m'}\Lambda_b
|j-1,m\rangle\langle j-1, m'| & \leftarrow \Gamma^{(5)}\\
&+D_z^{j,m}D_z^{j,m'}\Lambda_d
|j+1,m\rangle\langle j+1, m'|\bigr) & \leftarrow \Gamma^{(6)}
\end{align*}

(the coefficients related to the term $|j,m\rangle\langle j, m'|$
will be assigned ($\leftarrow$) to $\Gamma^{(1)}$ and so on.)

\begin{align*}
|\varphi_w|^2\sum_{n=1}^N J_-^{(n)}|j,m\rangle\langle j,m'|J_+^{(n)}
&= |\varphi_w|^2 \bigl(A_-^{j,m}A_-^{j,m'} \Lambda_a
|j,m-1\rangle\langle j,m'-1| & \leftarrow \Gamma^{(2)}\\ 
&+ B_-^{j,m}B_-^{j,m'}\Lambda_b
|j-1,m-1\rangle\langle j-1, m'-1| & \leftarrow \Gamma^{(3)} \\
&+D_-^{j,m}D_-^{j,m'}\Lambda_d
|j+1,m-1\rangle\langle j+1, m'-1|\bigr) & \leftarrow \Gamma^{(4)}
\end{align*}

\begin{align*}
|\varphi_w|^2\sum_{n=1}^N J_+^{(n)}|j,m\rangle\langle j,m'|J_-^{(n)}  
&=|\varphi_w|^2 \bigl(A_+^{j,m}A_+^{j,m'}\Lambda_a
 |j,m+1\rangle\langle j,m'+1| & \leftarrow \Gamma^{(8)} \\
&+ B_+^{j,m}B_+^{j,m'}\Lambda_b
|j-1,m+1\rangle\langle j-1, m'+1| & \leftarrow \Gamma^{(7)} \\
&+D_+^{j,m}D_+^{j,m'}\Lambda_d
|j+1,m+1\rangle\langle j+1, m'+1|\bigr) & \leftarrow \Gamma^{(9)}
\end{align*}
(note that $J_-^{(n)}$ becomes $J_+^{(n)\dagger}$ as in Eq. \eqref{eq:sum1})

\begin{align*}
(\varphi_w^*)^2\sum_{n=1}^N J_+^{(n)}|j,m\rangle\langle j,m'|J_+^{(n)}
&= (\varphi_w^*)^2 \bigl(A_+^{j,m}A_-^{j,m'} \Lambda_a
|j,m+1\rangle\langle j,m'-1| & \leftarrow \Gamma^{(10)}\\ 
&+ B_+^{j,m}B_-^{j,m'}\Lambda_b
|j-1,m+1\rangle\langle j-1, m'-1| & \leftarrow \Gamma^{(11)} \\
&+D_+^{j,m}D_-^{j,m'}\Lambda_d
|j+1,m+1\rangle\langle j+1, m'-1|\bigr) & \leftarrow \Gamma^{(12)}
\end{align*}

\begin{align*}
\varphi_w^*\varphi_z\sum_{n=1}^N J_+^{(n)}|j,m\rangle\langle j,m'|J_z^{(n)}
&= \varphi_w^*\varphi_z \bigl(A_+^{j,m}m' \Lambda_a
|j,m+1\rangle\langle j,m'| & \leftarrow \Gamma^{(13)}\\ 
&+ B_+^{j,m}B_z^{j,m'}\Lambda_b
|j-1,m+1\rangle\langle j-1, m'| & \leftarrow \Gamma^{(14)} \\
&+D_+^{j,m}D_z^{j,m'}\Lambda_d
|j+1,m+1\rangle\langle j+1, m'|\bigr) & \leftarrow \Gamma^{(15)}
\end{align*}

\begin{align*}
\varphi_w^2\sum_{n=1}^N J_-^{(n)}|j,m\rangle\langle j,m'|J_-^{(n)}
&= \varphi_w^2 \bigl(A_-^{j,m}A_+^{j,m'}\Lambda_a
 |j,m-1\rangle\langle j,m'+1| & \leftarrow \Gamma^{(16)}\\ 
&+ B_-^{j,m}B_+^{j,m'}\Lambda_b
|j-1,m-1\rangle\langle j-1, m' +1| & \leftarrow \Gamma^{(17)} \\
&+D_-^{j,m}D_+^{j,m'}\Lambda_d
|j+1,m-1\rangle\langle j+1, m'+1|\bigr) & \leftarrow \Gamma^{(18)}
\end{align*}

\begin{align*}
\varphi_w\varphi_z\sum_{n=1}^N J_-^{(n)}|j,m\rangle\langle j,m'|J_z^{(n)}
&= \varphi_w\varphi_z \bigl(A_-^{j,m}m' \Lambda_a
|j,m-1\rangle\langle j,m'| & \leftarrow \Gamma^{(19)}\\ 
&+ B_-^{j,m}B_z^{j,m'}\Lambda_b
|j-1,m-1\rangle\langle j-1, m'| & \leftarrow \Gamma^{(20)} \\
&+D_-^{j,m}D_z^{j,m'}\Lambda_d
|j+1,m-1\rangle\langle j+1, m'|\bigr) & \leftarrow \Gamma^{(21)}
\end{align*}

\begin{align*}
\varphi_w^*\varphi_z\sum_{n=1}^N J_z^{(n)}|j,m\rangle\langle j,m'|J_+^{(n)}
&= \varphi_w^*\varphi_z \bigl(m A_-^{j,m}\Lambda_a
|j,m\rangle\langle j,m'-1| & \leftarrow \Gamma^{(22)}\\ 
&+ B_z^{j,m}B_-^{j,m'}\Lambda_b
|j-1,m\rangle\langle j-1, m' -1| & \leftarrow \Gamma^{(23)} \\
&+D_z^{j,m}D_-^{j,m'}\Lambda_d
|j+1,m\rangle\langle j+1, m'-1|\bigr) & \leftarrow \Gamma^{(24)}
\end{align*}

\begin{align*}
\varphi_w\varphi_z\sum_{n=1}^N J_z^{(n)}|j,m\rangle\langle j,m'|J_-^{(n)}
&= \varphi_w\varphi_z \bigl(m A_+^{j,m}\Lambda_a
|j,m\rangle\langle j,m'+1| & \leftarrow \Gamma^{(25)}\\ 
&+ B_z^{j,m}B_+^{j,m'}\Lambda_b
|j-1,m\rangle\langle j-1, m' +1| & \leftarrow \Gamma^{(26)} \\
&+D_z^{j,m}D_+^{j,m'}\Lambda_d
|j+1,m\rangle\langle j+1, m'+1|\bigr) & \leftarrow \Gamma^{(27)}
\end{align*}

We collect all coefficients correspond to each $|\cdot,\cdot\rangle\langle\cdot,\cdot|$
and assign as $\Gamma^{(i)}$, where $i = 1, \cdots, 27$ as following:

\begin{center}
\small
\begin{tabular}{ l |c |c |c |c |c |c |c |c |c }
 \backslashbox{$m$}{$m'$} & $\langle j-1,m'-1|$ & $\langle j-1,m'|$ & $\langle j-1,m'+1|$ &$\langle j,m'-1|$
  & $\langle j,m'|$ & $\langle j,m'+1|$ &$\langle j+1,m'-1|$ & $\langle j+1,m'|$ & $\langle j+1,m'+1|$\\ 
  \hline
 $|j-1,m-1\rangle$ & $\Gamma^{(3)}$ & $\Gamma^{(20)}$ & $\Gamma^{(17)}$ &&&&&&\\ 
   \hline
 $|j-1,m\rangle$ & $\Gamma^{(23)}$ & $\Gamma^{(5)}$ & $\Gamma^{(26)}$ &&&&&&\\ 
    \hline
 $|j-1,m+1\rangle$ & $\Gamma^{(11)}$ & $\Gamma^{(14)}$ & $\Gamma^{(7)}$ &&&&&&\\ 
    \hline
 $|j,m-1\rangle$ &&&& $\Gamma^{(2)}$ & $\Gamma^{(19)}$ & $\Gamma^{(16)}$ &&&\\   
     \hline
 $|j,m\rangle$ &&&& $\Gamma^{(22)}$ & $\Gamma^{(1)}$ & $\Gamma^{(25)}$ &&&\\  
     \hline
 $|j,m+1\rangle$ &&&& $\Gamma^{(10)}$ & $\Gamma^{(13)}$ & $\Gamma^{(8)}$ &&&\\  
    \hline
 $|j+1,m-1\rangle$ &&&&&&& $\Gamma^{(4)}$ & $\Gamma^{(21)}$ & $\Gamma^{(18)}$\\  
   \hline
 $|j+1,m\rangle$ &&&&&&& $\Gamma^{(24)}$ & $\Gamma^{(6)}$ & $\Gamma^{(27)}$\\  
    \hline
 $|j+1,m+1\rangle$ &&&&&&& $\Gamma^{(12)}$ & $\Gamma^{(15)}$ & $\Gamma^{(9)}$\\   
\end{tabular}
\end{center}

Explicitly, we have
\begin{center}
\begin{tabular}{ l|l|l }
$ \Gamma^{(1)} = 2\gamma_t
(\varphi_z^2mm'\Lambda_a-N)
$
&
$ \Gamma^{(10)} = 2\gamma_t
(\varphi^*_w)^2 A_+^{j,m}A_-^{j,m'}\Lambda_a
$
&
$ \Gamma^{(19)} = 2\gamma_t
\varphi_w \varphi_z A_-^{j,m}m'\Lambda_a
$\\
$ \Gamma^{(2)} = 2\gamma_t
|\varphi_w|^2A_-^{j,m}A_-^{j,m'}
\Lambda_a
$
&
$ \Gamma^{(11)} = 2\gamma_t
(\varphi^*_w)^2 B_+^{j,m}B_-^{j,m'}\Lambda_b
$
&
$ \Gamma^{(20)} = 2\gamma_t
\varphi_w \varphi_z B_-^{j,m}B_z^{j,m'}\Lambda_b
$\\
$ \Gamma^{(3)} = 2\gamma_t
|\varphi_w|^2B_-^{j,m}B_-^{j,m'}
\Lambda_b
$
&
$ \Gamma^{(12)} = 2\gamma_t
(\varphi^*_w)^2 D_+^{j,m}D_-^{j,m'}\Lambda_d
$
&
$ \Gamma^{(21)} = 2\gamma_t
\varphi_w \varphi_z D_-^{j,m}D_z^{j,m'}\Lambda_d
$\\
$ \Gamma^{(4)} = 2\gamma_t
|\varphi_w|^2D_-^{j,m}D_-^{j,m'}
\Lambda_d
$
&
$ \Gamma^{(13)} = 2\gamma_t
\varphi^*_w\varphi_z A_+^{j,m}m'\Lambda_a
$
&
$ \Gamma^{(22)} = 2\gamma_t
\varphi_w^* \varphi_z mA_-^{j,m'}\Lambda_a
$\\
$ \Gamma^{(5)} = 2\gamma_t
\varphi_z^2B_z^{j,m}B_z^{j,m'}
\Lambda_b
$
&
$ \Gamma^{(14)} = 2\gamma_t
\varphi^*_w\varphi_z B_+^{j,m}B_z^{j,m'}\Lambda_b
$
&
$ \Gamma^{(23)} = 2\gamma_t
\varphi_w^* \varphi_z B_z^{j,m}B_-^{j,m'}\Lambda_b
$\\
$ \Gamma^{(6)} = 2\gamma_t
\varphi_z^2D_z^{j,m}D_z^{j,m'}
\Lambda_d
$
&
$ \Gamma^{(15)} = 2\gamma_t
\varphi^*_w\varphi_z D_+^{j,m}D_z^{j,m'}\Lambda_d
$
&
$ \Gamma^{(24)} = 2\gamma_t
\varphi_w^* \varphi_z D_z^{j,m}D_-^{j,m'}\Lambda_d
$\\
$ \Gamma^{(7)} = 2\gamma_t
|\varphi_w|^2B_+^{j,m}B_+^{j,m'}
\Lambda_b
$
&
$ \Gamma^{(16)} = 2\gamma_t
\varphi_w^2 A_-^{j,m}A_+^{j,m'}\Lambda_a
$
&
$ \Gamma^{(25)} = 2\gamma_t
\varphi_w \varphi_z mA_+^{j,m'}\Lambda_a
$\\
$ \Gamma^{(8)} = 2\gamma_t
|\varphi_w|^2A_+^{j,m}A_+^{j,m'}
\Lambda_a
$
&
$ \Gamma^{(17)} = 2\gamma_t
\varphi_w^2 B_-^{j,m}B_+^{j,m'}\Lambda_b
$
&
$ \Gamma^{(26)} = 2\gamma_t
\varphi_w \varphi_z B_z^{j,m}B_+^{j,m'}\Lambda_b
$\\
$ \Gamma^{(9)} = 2\gamma_t
|\varphi_w|^2D_+^{j,m}D_+^{j,m'}
\Lambda_d
$
&
$ \Gamma^{(18)} = 2\gamma_t
\varphi_w^2 D_-^{j,m}D_+^{j,m'}\Lambda_d
$
&
$ \Gamma^{(27)} = 2\gamma_t
\varphi_w \varphi_z D_z^{j,m}D_+^{j,m'}\Lambda_d
$\\
\end{tabular}
\end{center}
\begin{figure} [t!]
\includegraphics[width=12cm]{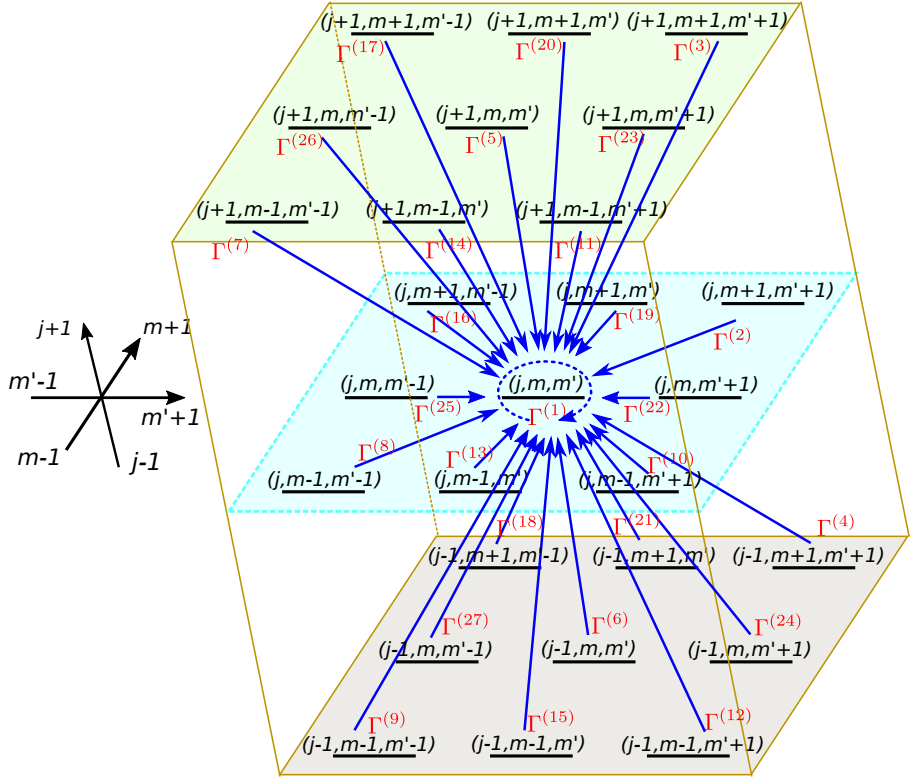}
\caption{(Color online)
Sketch of the dynamics coupling given by a Dicke state,
represented in terms of the coefficients $\Gamma^{(i)}$.
We show the action of each coefficient onto a given Dicke state. 
All processes contribute to the coefficient $\Gamma^{(1)}$.
}
\label{fig:coupling_Gamma}
\end{figure} 
thus the equation can be solved. 
In the numerical calculation, 
we have extended the Permutational-Invariant Quantum Solver (PIQS) 
library in QutiP \cite{PhysRevA.98.063815} using our analysis in this appendix.

Finally, $\rho_t(\bm\phi)$ is given by the evolution
$U(\bm\phi)\rho_t U^{\dagger}(\bm\phi)$.

\setcounter{equation}{0}
\renewcommand{\theequation}{C.\arabic{equation}}
\section{Concrete calculation of the QFIM}\label{appC}
For concreteness, to calculate the QFIM, we first derive 
the term ${\partial \rho_t(\bm\phi)}/{\partial \phi_k}$ as
\begin{align}\label{eq:der_psi_phi}
\notag \partial_k \rho_t(\bm\phi)&= 
\partial_k \bigl[U(\bm\phi)\rho_tU^\dagger(\bm\phi)\bigr]\\
&=\partial_k U(\bm\phi)
\rho_tU^\dagger(\bm\phi)
+U(\bm\phi)\rho_t \partial_k U^\dagger(\bm\phi),
\end{align}
where $U(\bm\phi) = e^{-itH(\bm\phi)}$.
Detailed calculation \cite{doi:10.1063/1.1705306}:
\begin{align}\label{eq:der_psi_phi_temp}
\notag \partial_k U(\bm\phi)
&= \partial_k e^{-itH(\bm\phi)}\\
\notag &= -i\int_0^t due^{-i(t-u)H(\bm\phi)}
[\partial_k H(\bm\phi)]e^{-iuH(\bm\phi)}\\
\notag &= -ie^{-itH(\bm\phi)}\int_0^t due^{iuH(\bm\phi)}J_k e^{-iuH(\bm\phi)}
\\
&=-iU(\bm\phi)A_k,
\end{align}
where
\begin{align}\label{eq:Aalpha}
A_k = \int_0^t du\ e^{iu H(\bm\phi)}J_k
e^{-iu H(\bm\phi)},
\end{align}
is a Hermitian operator \cite{doi:10.1063/1.1705306,PhysRevA.90.022117,PhysRevLett.116.030801}.
To solve Eq.~\eqref{eq:Aalpha}, we follow the method
described in Refs.~\cite{doi:10.1063/1.1705306,PhysRevA.90.022117},
and we use this form to calculate QFIM.
Therein, for $t \ll 1$, we have
\begin{align}\label{eq:Atsmall}
A_k \approx tJ_k.
\end{align}
For arbitrary large $t$, we have \cite{doi:10.1063/1.1705306,PhysRevA.90.022117}
\begin{align}\label{eq:Atlarge}
A_k = t\sum_{\{l|\lambda_l=0\}}{\rm Tr}
[\Gamma_l^\dagger J_k]\Gamma_l
-i\sum_{\{l|\lambda_l\ne0\}}
\frac{1-e^{-i\lambda_lt}}{\lambda_l}
{\rm Tr}[\Gamma_l^\dagger J_k]\Gamma_l,
\end{align}
where $\Gamma$ satisfies the eigenvalue 
equation:
\begin{align}\label{eq:mathH}
\mathcal {H}(\bm \phi)\Gamma 
\equiv [H(\bm\phi), \Gamma]
= \lambda\Gamma.
\end{align}
$\mathcal {H}(\bm \phi)$ is a Hermitian superoperator
of $H(\bm \phi)$, which has $d_D^2$ real eigenvalues: 
$\lambda_1, \cdots, \lambda_{d_D^2}$. 
We also denote $\Gamma_l, l = 1, \cdots, d_D^2$
are orthonormal eigenvalues of $\Gamma$.

Finally, substituting Eq.~\eqref{eq:der_psi_phi} into 
Eq.~\eqref{eq:Qsol_rec} in the main text, 
we obtain the QFIM.

\end{widetext}

\bibliography{refs}

\begin{thebibliography}{69}%
\makeatletter
\providecommand \@ifxundefined [1]{%
 \@ifx{#1\undefined}
}%
\providecommand \@ifnum [1]{%
 \ifnum #1\expandafter \@firstoftwo
 \else \expandafter \@secondoftwo
 \fi
}%
\providecommand \@ifx [1]{%
 \ifx #1\expandafter \@firstoftwo
 \else \expandafter \@secondoftwo
 \fi
}%
\providecommand \natexlab [1]{#1}%
\providecommand \enquote  [1]{``#1''}%
\providecommand \bibnamefont  [1]{#1}%
\providecommand \bibfnamefont [1]{#1}%
\providecommand \citenamefont [1]{#1}%
\providecommand \href@noop [0]{\@secondoftwo}%
\providecommand \href [0]{\begingroup \@sanitize@url \@href}%
\providecommand \@href[1]{\@@startlink{#1}\@@href}%
\providecommand \@@href[1]{\endgroup#1\@@endlink}%
\providecommand \@sanitize@url [0]{\catcode `\\12\catcode `\$12\catcode
  `\&12\catcode `\#12\catcode `\^12\catcode `\_12\catcode `\%12\relax}%
\providecommand \@@startlink[1]{}%
\providecommand \@@endlink[0]{}%
\providecommand \url  [0]{\begingroup\@sanitize@url \@url }%
\providecommand \@url [1]{\endgroup\@href {#1}{\urlprefix }}%
\providecommand \urlprefix  [0]{URL }%
\providecommand \Eprint [0]{\href }%
\providecommand \doibase [0]{http://dx.doi.org/}%
\providecommand \selectlanguage [0]{\@gobble}%
\providecommand \bibinfo  [0]{\@secondoftwo}%
\providecommand \bibfield  [0]{\@secondoftwo}%
\providecommand \translation [1]{[#1]}%
\providecommand \BibitemOpen [0]{}%
\providecommand \bibitemStop [0]{}%
\providecommand \bibitemNoStop [0]{.\EOS\space}%
\providecommand \EOS [0]{\spacefactor3000\relax}%
\providecommand \BibitemShut  [1]{\csname bibitem#1\endcsname}%
\let\auto@bib@innerbib\@empty
\bibitem [{\citenamefont {Pezz\'e}\ and\ \citenamefont
  {Smerzi}(2009)}]{PhysRevLett.102.100401}%
  \BibitemOpen
  \bibfield  {author} {\bibinfo {author} {\bibfnamefont {L.}~\bibnamefont
  {Pezz\'e}}\ and\ \bibinfo {author} {\bibfnamefont {A.}~\bibnamefont
  {Smerzi}},\ }\href {\doibase 10.1103/PhysRevLett.102.100401} {\bibfield
  {journal} {\bibinfo  {journal} {Phys. Rev. Lett.}\ }\textbf {\bibinfo
  {volume} {102}},\ \bibinfo {pages} {100401} (\bibinfo {year}
  {2009})}\BibitemShut {NoStop}%
\bibitem [{\citenamefont {Huelga}\ \emph {et~al.}(1997)\citenamefont {Huelga},
  \citenamefont {Macchiavello}, \citenamefont {Pellizzari}, \citenamefont
  {Ekert}, \citenamefont {Plenio},\ and\ \citenamefont
  {Cirac}}]{PhysRevLett.79.3865}%
  \BibitemOpen
  \bibfield  {author} {\bibinfo {author} {\bibfnamefont {S.~F.}\ \bibnamefont
  {Huelga}}, \bibinfo {author} {\bibfnamefont {C.}~\bibnamefont
  {Macchiavello}}, \bibinfo {author} {\bibfnamefont {T.}~\bibnamefont
  {Pellizzari}}, \bibinfo {author} {\bibfnamefont {A.~K.}\ \bibnamefont
  {Ekert}}, \bibinfo {author} {\bibfnamefont {M.~B.}\ \bibnamefont {Plenio}}, \
  and\ \bibinfo {author} {\bibfnamefont {J.~I.}\ \bibnamefont {Cirac}},\ }\href
  {\doibase 10.1103/PhysRevLett.79.3865} {\bibfield  {journal} {\bibinfo
  {journal} {Phys. Rev. Lett.}\ }\textbf {\bibinfo {volume} {79}},\ \bibinfo
  {pages} {3865} (\bibinfo {year} {1997})}\BibitemShut {NoStop}%
\bibitem [{\citenamefont {Wineland}\ \emph {et~al.}(1992)\citenamefont
  {Wineland}, \citenamefont {Bollinger}, \citenamefont {Itano}, \citenamefont
  {Moore},\ and\ \citenamefont {Heinzen}}]{PhysRevA.46.R6797}%
  \BibitemOpen
  \bibfield  {author} {\bibinfo {author} {\bibfnamefont {D.~J.}\ \bibnamefont
  {Wineland}}, \bibinfo {author} {\bibfnamefont {J.~J.}\ \bibnamefont
  {Bollinger}}, \bibinfo {author} {\bibfnamefont {W.~M.}\ \bibnamefont
  {Itano}}, \bibinfo {author} {\bibfnamefont {F.~L.}\ \bibnamefont {Moore}}, \
  and\ \bibinfo {author} {\bibfnamefont {D.~J.}\ \bibnamefont {Heinzen}},\
  }\href {\doibase 10.1103/PhysRevA.46.R6797} {\bibfield  {journal} {\bibinfo
  {journal} {Phys. Rev. A}\ }\textbf {\bibinfo {volume} {46}},\ \bibinfo
  {pages} {R6797} (\bibinfo {year} {1992})}\BibitemShut {NoStop}%
\bibitem [{\citenamefont {Wineland}\ \emph {et~al.}(1994)\citenamefont
  {Wineland}, \citenamefont {Bollinger}, \citenamefont {Itano},\ and\
  \citenamefont {Heinzen}}]{PhysRevA.50.67}%
  \BibitemOpen
  \bibfield  {author} {\bibinfo {author} {\bibfnamefont {D.~J.}\ \bibnamefont
  {Wineland}}, \bibinfo {author} {\bibfnamefont {J.~J.}\ \bibnamefont
  {Bollinger}}, \bibinfo {author} {\bibfnamefont {W.~M.}\ \bibnamefont
  {Itano}}, \ and\ \bibinfo {author} {\bibfnamefont {D.~J.}\ \bibnamefont
  {Heinzen}},\ }\href {\doibase 10.1103/PhysRevA.50.67} {\bibfield  {journal}
  {\bibinfo  {journal} {Phys. Rev. A}\ }\textbf {\bibinfo {volume} {50}},\
  \bibinfo {pages} {67} (\bibinfo {year} {1994})}\BibitemShut {NoStop}%
\bibitem [{\citenamefont {Giovannetti}\ \emph {et~al.}(2004)\citenamefont
  {Giovannetti}, \citenamefont {Lloyd},\ and\ \citenamefont
  {Maccone}}]{Giovannetti1330}%
  \BibitemOpen
  \bibfield  {author} {\bibinfo {author} {\bibfnamefont {V.}~\bibnamefont
  {Giovannetti}}, \bibinfo {author} {\bibfnamefont {S.}~\bibnamefont {Lloyd}},
  \ and\ \bibinfo {author} {\bibfnamefont {L.}~\bibnamefont {Maccone}},\ }\href
  {\doibase 10.1126/science.1104149} {\bibfield  {journal} {\bibinfo  {journal}
  {Science}\ }\textbf {\bibinfo {volume} {306}},\ \bibinfo {pages} {1330}
  (\bibinfo {year} {2004})},\ \Eprint
  {http://arxiv.org/abs/https://science.sciencemag.org/content/306/5700/1330.full.pdf}
  {https://science.sciencemag.org/content/306/5700/1330.full.pdf} \BibitemShut
  {NoStop}%
\bibitem [{\citenamefont {Giovannetti}\ \emph {et~al.}(2006)\citenamefont
  {Giovannetti}, \citenamefont {Lloyd},\ and\ \citenamefont
  {Maccone}}]{PhysRevLett.96.010401}%
  \BibitemOpen
  \bibfield  {author} {\bibinfo {author} {\bibfnamefont {V.}~\bibnamefont
  {Giovannetti}}, \bibinfo {author} {\bibfnamefont {S.}~\bibnamefont {Lloyd}},
  \ and\ \bibinfo {author} {\bibfnamefont {L.}~\bibnamefont {Maccone}},\ }\href
  {\doibase 10.1103/PhysRevLett.96.010401} {\bibfield  {journal} {\bibinfo
  {journal} {Phys. Rev. Lett.}\ }\textbf {\bibinfo {volume} {96}},\ \bibinfo
  {pages} {010401} (\bibinfo {year} {2006})}\BibitemShut {NoStop}%
\bibitem [{\citenamefont {Jones}\ \emph {et~al.}(2009)\citenamefont {Jones},
  \citenamefont {Karlen}, \citenamefont {Fitzsimons}, \citenamefont {Ardavan},
  \citenamefont {Benjamin}, \citenamefont {Briggs},\ and\ \citenamefont
  {Morton}}]{Jones1166}%
  \BibitemOpen
  \bibfield  {author} {\bibinfo {author} {\bibfnamefont {J.~A.}\ \bibnamefont
  {Jones}}, \bibinfo {author} {\bibfnamefont {S.~D.}\ \bibnamefont {Karlen}},
  \bibinfo {author} {\bibfnamefont {J.}~\bibnamefont {Fitzsimons}}, \bibinfo
  {author} {\bibfnamefont {A.}~\bibnamefont {Ardavan}}, \bibinfo {author}
  {\bibfnamefont {S.~C.}\ \bibnamefont {Benjamin}}, \bibinfo {author}
  {\bibfnamefont {G.~A.~D.}\ \bibnamefont {Briggs}}, \ and\ \bibinfo {author}
  {\bibfnamefont {J.~J.~L.}\ \bibnamefont {Morton}},\ }\href {\doibase
  10.1126/science.1170730} {\bibfield  {journal} {\bibinfo  {journal}
  {Science}\ }\textbf {\bibinfo {volume} {324}},\ \bibinfo {pages} {1166}
  (\bibinfo {year} {2009})},\ \Eprint
  {http://arxiv.org/abs/https://science.sciencemag.org/content/324/5931/1166.full.pdf}
  {https://science.sciencemag.org/content/324/5931/1166.full.pdf} \BibitemShut
  {NoStop}%
\bibitem [{\citenamefont {Simmons}\ \emph {et~al.}(2010)\citenamefont
  {Simmons}, \citenamefont {Jones}, \citenamefont {Karlen}, \citenamefont
  {Ardavan},\ and\ \citenamefont {Morton}}]{PhysRevA.82.022330}%
  \BibitemOpen
  \bibfield  {author} {\bibinfo {author} {\bibfnamefont {S.}~\bibnamefont
  {Simmons}}, \bibinfo {author} {\bibfnamefont {J.~A.}\ \bibnamefont {Jones}},
  \bibinfo {author} {\bibfnamefont {S.~D.}\ \bibnamefont {Karlen}}, \bibinfo
  {author} {\bibfnamefont {A.}~\bibnamefont {Ardavan}}, \ and\ \bibinfo
  {author} {\bibfnamefont {J.~J.~L.}\ \bibnamefont {Morton}},\ }\href {\doibase
  10.1103/PhysRevA.82.022330} {\bibfield  {journal} {\bibinfo  {journal} {Phys.
  Rev. A}\ }\textbf {\bibinfo {volume} {82}},\ \bibinfo {pages} {022330}
  (\bibinfo {year} {2010})}\BibitemShut {NoStop}%
\bibitem [{\citenamefont {Zaiser}\ \emph {et~al.}(2016)\citenamefont {Zaiser},
  \citenamefont {Rendler}, \citenamefont {Jakobi}, \citenamefont {Wolf},
  \citenamefont {Lee}, \citenamefont {Wagner}, \citenamefont {Bergholm},
  \citenamefont {Schulte-Herbr{\"u}ggen}, \citenamefont {Neumann},\ and\
  \citenamefont {Wrachtrup}}]{Zaiser2016}%
  \BibitemOpen
  \bibfield  {author} {\bibinfo {author} {\bibfnamefont {S.}~\bibnamefont
  {Zaiser}}, \bibinfo {author} {\bibfnamefont {T.}~\bibnamefont {Rendler}},
  \bibinfo {author} {\bibfnamefont {I.}~\bibnamefont {Jakobi}}, \bibinfo
  {author} {\bibfnamefont {T.}~\bibnamefont {Wolf}}, \bibinfo {author}
  {\bibfnamefont {S.-Y.}\ \bibnamefont {Lee}}, \bibinfo {author} {\bibfnamefont
  {S.}~\bibnamefont {Wagner}}, \bibinfo {author} {\bibfnamefont
  {V.}~\bibnamefont {Bergholm}}, \bibinfo {author} {\bibfnamefont
  {T.}~\bibnamefont {Schulte-Herbr{\"u}ggen}}, \bibinfo {author} {\bibfnamefont
  {P.}~\bibnamefont {Neumann}}, \ and\ \bibinfo {author} {\bibfnamefont
  {J.}~\bibnamefont {Wrachtrup}},\ }\href {\doibase 10.1038/ncomms12279}
  {\bibfield  {journal} {\bibinfo  {journal} {Nature Communications}\ }\textbf
  {\bibinfo {volume} {7}},\ \bibinfo {pages} {12279} (\bibinfo {year}
  {2016})}\BibitemShut {NoStop}%
\bibitem [{\citenamefont {Matsuzaki}\ \emph {et~al.}(2018)\citenamefont
  {Matsuzaki}, \citenamefont {Benjamin}, \citenamefont {Nakayama},
  \citenamefont {Saito},\ and\ \citenamefont {Munro}}]{PhysRevLett.120.140501}%
  \BibitemOpen
  \bibfield  {author} {\bibinfo {author} {\bibfnamefont {Y.}~\bibnamefont
  {Matsuzaki}}, \bibinfo {author} {\bibfnamefont {S.}~\bibnamefont {Benjamin}},
  \bibinfo {author} {\bibfnamefont {S.}~\bibnamefont {Nakayama}}, \bibinfo
  {author} {\bibfnamefont {S.}~\bibnamefont {Saito}}, \ and\ \bibinfo {author}
  {\bibfnamefont {W.~J.}\ \bibnamefont {Munro}},\ }\href {\doibase
  10.1103/PhysRevLett.120.140501} {\bibfield  {journal} {\bibinfo  {journal}
  {Phys. Rev. Lett.}\ }\textbf {\bibinfo {volume} {120}},\ \bibinfo {pages}
  {140501} (\bibinfo {year} {2018})}\BibitemShut {NoStop}%
\bibitem [{\citenamefont {Zhang}\ \emph {et~al.}(2015)\citenamefont {Zhang},
  \citenamefont {Datta},\ and\ \citenamefont
  {Walmsley}}]{PhysRevLett.114.210801}%
  \BibitemOpen
  \bibfield  {author} {\bibinfo {author} {\bibfnamefont {L.}~\bibnamefont
  {Zhang}}, \bibinfo {author} {\bibfnamefont {A.}~\bibnamefont {Datta}}, \ and\
  \bibinfo {author} {\bibfnamefont {I.~A.}\ \bibnamefont {Walmsley}},\ }\href
  {\doibase 10.1103/PhysRevLett.114.210801} {\bibfield  {journal} {\bibinfo
  {journal} {Phys. Rev. Lett.}\ }\textbf {\bibinfo {volume} {114}},\ \bibinfo
  {pages} {210801} (\bibinfo {year} {2015})}\BibitemShut {NoStop}%
\bibitem [{\citenamefont {Pang}\ \emph {et~al.}(2014)\citenamefont {Pang},
  \citenamefont {Dressel},\ and\ \citenamefont
  {Brun}}]{PhysRevLett.113.030401}%
  \BibitemOpen
  \bibfield  {author} {\bibinfo {author} {\bibfnamefont {S.}~\bibnamefont
  {Pang}}, \bibinfo {author} {\bibfnamefont {J.}~\bibnamefont {Dressel}}, \
  and\ \bibinfo {author} {\bibfnamefont {T.~A.}\ \bibnamefont {Brun}},\ }\href
  {\doibase 10.1103/PhysRevLett.113.030401} {\bibfield  {journal} {\bibinfo
  {journal} {Phys. Rev. Lett.}\ }\textbf {\bibinfo {volume} {113}},\ \bibinfo
  {pages} {030401} (\bibinfo {year} {2014})}\BibitemShut {NoStop}%
\bibitem [{\citenamefont {Pang}\ and\ \citenamefont
  {Brun}(2015{\natexlab{a}})}]{PhysRevA.92.012120}%
  \BibitemOpen
  \bibfield  {author} {\bibinfo {author} {\bibfnamefont {S.}~\bibnamefont
  {Pang}}\ and\ \bibinfo {author} {\bibfnamefont {T.~A.}\ \bibnamefont
  {Brun}},\ }\href {\doibase 10.1103/PhysRevA.92.012120} {\bibfield  {journal}
  {\bibinfo  {journal} {Phys. Rev. A}\ }\textbf {\bibinfo {volume} {92}},\
  \bibinfo {pages} {012120} (\bibinfo {year} {2015}{\natexlab{a}})}\BibitemShut
  {NoStop}%
\bibitem [{\citenamefont {Pang}\ and\ \citenamefont
  {Brun}(2015{\natexlab{b}})}]{PhysRevLett.115.120401}%
  \BibitemOpen
  \bibfield  {author} {\bibinfo {author} {\bibfnamefont {S.}~\bibnamefont
  {Pang}}\ and\ \bibinfo {author} {\bibfnamefont {T.~A.}\ \bibnamefont
  {Brun}},\ }\href {\doibase 10.1103/PhysRevLett.115.120401} {\bibfield
  {journal} {\bibinfo  {journal} {Phys. Rev. Lett.}\ }\textbf {\bibinfo
  {volume} {115}},\ \bibinfo {pages} {120401} (\bibinfo {year}
  {2015}{\natexlab{b}})}\BibitemShut {NoStop}%
\bibitem [{\citenamefont {Jordan}\ \emph {et~al.}(2015)\citenamefont {Jordan},
  \citenamefont {Tollaksen}, \citenamefont {Troupe}, \citenamefont {Dressel},\
  and\ \citenamefont {Aharonov}}]{Jordan2015}%
  \BibitemOpen
  \bibfield  {author} {\bibinfo {author} {\bibfnamefont {A.~N.}\ \bibnamefont
  {Jordan}}, \bibinfo {author} {\bibfnamefont {J.}~\bibnamefont {Tollaksen}},
  \bibinfo {author} {\bibfnamefont {J.~E.}\ \bibnamefont {Troupe}}, \bibinfo
  {author} {\bibfnamefont {J.}~\bibnamefont {Dressel}}, \ and\ \bibinfo
  {author} {\bibfnamefont {Y.}~\bibnamefont {Aharonov}},\ }\href {\doibase
  10.1007/s40509-015-0036-8} {\bibfield  {journal} {\bibinfo  {journal}
  {Quantum Studies: Mathematics and Foundations}\ }\textbf {\bibinfo {volume}
  {2}},\ \bibinfo {pages} {5} (\bibinfo {year} {2015})}\BibitemShut {NoStop}%
\bibitem [{\citenamefont {Ho}\ and\ \citenamefont {Kondo}(2019)}]{HO2019153}%
  \BibitemOpen
  \bibfield  {author} {\bibinfo {author} {\bibfnamefont {L.~B.}\ \bibnamefont
  {Ho}}\ and\ \bibinfo {author} {\bibfnamefont {Y.}~\bibnamefont {Kondo}},\
  }\href {\doibase https://doi.org/10.1016/j.physleta.2018.10.041} {\bibfield
  {journal} {\bibinfo  {journal} {Physics Letters A}\ }\textbf {\bibinfo
  {volume} {383}},\ \bibinfo {pages} {153 } (\bibinfo {year}
  {2019})}\BibitemShut {NoStop}%
\bibitem [{\citenamefont {Zhou}\ \emph {et~al.}(2018)\citenamefont {Zhou},
  \citenamefont {Zhang}, \citenamefont {Preskill},\ and\ \citenamefont
  {Jiang}}]{Zhou2018}%
  \BibitemOpen
  \bibfield  {author} {\bibinfo {author} {\bibfnamefont {S.}~\bibnamefont
  {Zhou}}, \bibinfo {author} {\bibfnamefont {M.}~\bibnamefont {Zhang}},
  \bibinfo {author} {\bibfnamefont {J.}~\bibnamefont {Preskill}}, \ and\
  \bibinfo {author} {\bibfnamefont {L.}~\bibnamefont {Jiang}},\ }\href
  {\doibase 10.1038/s41467-017-02510-3} {\bibfield  {journal} {\bibinfo
  {journal} {Nature Communications}\ }\textbf {\bibinfo {volume} {9}},\
  \bibinfo {pages} {78} (\bibinfo {year} {2018})}\BibitemShut {NoStop}%
\bibitem [{\citenamefont {Matsuzaki}\ \emph {et~al.}(2011)\citenamefont
  {Matsuzaki}, \citenamefont {Benjamin},\ and\ \citenamefont
  {Fitzsimons}}]{PhysRevA.84.012103}%
  \BibitemOpen
  \bibfield  {author} {\bibinfo {author} {\bibfnamefont {Y.}~\bibnamefont
  {Matsuzaki}}, \bibinfo {author} {\bibfnamefont {S.~C.}\ \bibnamefont
  {Benjamin}}, \ and\ \bibinfo {author} {\bibfnamefont {J.}~\bibnamefont
  {Fitzsimons}},\ }\href {\doibase 10.1103/PhysRevA.84.012103} {\bibfield
  {journal} {\bibinfo  {journal} {Phys. Rev. A}\ }\textbf {\bibinfo {volume}
  {84}},\ \bibinfo {pages} {012103} (\bibinfo {year} {2011})}\BibitemShut
  {NoStop}%
\bibitem [{\citenamefont {D\"ur}\ \emph {et~al.}(2014)\citenamefont {D\"ur},
  \citenamefont {Skotiniotis}, \citenamefont {Fr\"owis},\ and\ \citenamefont
  {Kraus}}]{PhysRevLett.112.080801}%
  \BibitemOpen
  \bibfield  {author} {\bibinfo {author} {\bibfnamefont {W.}~\bibnamefont
  {D\"ur}}, \bibinfo {author} {\bibfnamefont {M.}~\bibnamefont {Skotiniotis}},
  \bibinfo {author} {\bibfnamefont {F.}~\bibnamefont {Fr\"owis}}, \ and\
  \bibinfo {author} {\bibfnamefont {B.}~\bibnamefont {Kraus}},\ }\href
  {\doibase 10.1103/PhysRevLett.112.080801} {\bibfield  {journal} {\bibinfo
  {journal} {Phys. Rev. Lett.}\ }\textbf {\bibinfo {volume} {112}},\ \bibinfo
  {pages} {080801} (\bibinfo {year} {2014})}\BibitemShut {NoStop}%
\bibitem [{\citenamefont {Herrera-Mart\'{\i}}\ \emph
  {et~al.}(2015)\citenamefont {Herrera-Mart\'{\i}}, \citenamefont {Gefen},
  \citenamefont {Aharonov}, \citenamefont {Katz},\ and\ \citenamefont
  {Retzker}}]{PhysRevLett.115.200501}%
  \BibitemOpen
  \bibfield  {author} {\bibinfo {author} {\bibfnamefont {D.~A.}\ \bibnamefont
  {Herrera-Mart\'{\i}}}, \bibinfo {author} {\bibfnamefont {T.}~\bibnamefont
  {Gefen}}, \bibinfo {author} {\bibfnamefont {D.}~\bibnamefont {Aharonov}},
  \bibinfo {author} {\bibfnamefont {N.}~\bibnamefont {Katz}}, \ and\ \bibinfo
  {author} {\bibfnamefont {A.}~\bibnamefont {Retzker}},\ }\href {\doibase
  10.1103/PhysRevLett.115.200501} {\bibfield  {journal} {\bibinfo  {journal}
  {Phys. Rev. Lett.}\ }\textbf {\bibinfo {volume} {115}},\ \bibinfo {pages}
  {200501} (\bibinfo {year} {2015})}\BibitemShut {NoStop}%
\bibitem [{\citenamefont {Arrad}\ \emph {et~al.}(2014)\citenamefont {Arrad},
  \citenamefont {Vinkler}, \citenamefont {Aharonov},\ and\ \citenamefont
  {Retzker}}]{PhysRevLett.112.150801}%
  \BibitemOpen
  \bibfield  {author} {\bibinfo {author} {\bibfnamefont {G.}~\bibnamefont
  {Arrad}}, \bibinfo {author} {\bibfnamefont {Y.}~\bibnamefont {Vinkler}},
  \bibinfo {author} {\bibfnamefont {D.}~\bibnamefont {Aharonov}}, \ and\
  \bibinfo {author} {\bibfnamefont {A.}~\bibnamefont {Retzker}},\ }\href
  {\doibase 10.1103/PhysRevLett.112.150801} {\bibfield  {journal} {\bibinfo
  {journal} {Phys. Rev. Lett.}\ }\textbf {\bibinfo {volume} {112}},\ \bibinfo
  {pages} {150801} (\bibinfo {year} {2014})}\BibitemShut {NoStop}%
\bibitem [{\citenamefont {Kessler}\ \emph {et~al.}(2014)\citenamefont
  {Kessler}, \citenamefont {Lovchinsky}, \citenamefont {Sushkov},\ and\
  \citenamefont {Lukin}}]{PhysRevLett.112.150802}%
  \BibitemOpen
  \bibfield  {author} {\bibinfo {author} {\bibfnamefont {E.~M.}\ \bibnamefont
  {Kessler}}, \bibinfo {author} {\bibfnamefont {I.}~\bibnamefont {Lovchinsky}},
  \bibinfo {author} {\bibfnamefont {A.~O.}\ \bibnamefont {Sushkov}}, \ and\
  \bibinfo {author} {\bibfnamefont {M.~D.}\ \bibnamefont {Lukin}},\ }\href
  {\doibase 10.1103/PhysRevLett.112.150802} {\bibfield  {journal} {\bibinfo
  {journal} {Phys. Rev. Lett.}\ }\textbf {\bibinfo {volume} {112}},\ \bibinfo
  {pages} {150802} (\bibinfo {year} {2014})}\BibitemShut {NoStop}%
\bibitem [{\citenamefont {Matsuzaki}\ and\ \citenamefont
  {Benjamin}(2017)}]{matsuzaki2017magnetic}%
  \BibitemOpen
  \bibfield  {author} {\bibinfo {author} {\bibfnamefont {Y.}~\bibnamefont
  {Matsuzaki}}\ and\ \bibinfo {author} {\bibfnamefont {S.}~\bibnamefont
  {Benjamin}},\ }\href@noop {} {\bibfield  {journal} {\bibinfo  {journal}
  {Physical Review A}\ }\textbf {\bibinfo {volume} {95}},\ \bibinfo {pages}
  {032303} (\bibinfo {year} {2017})}\BibitemShut {NoStop}%
\bibitem [{\citenamefont {Demkowicz-Dobrza\ifmmode~\acute{n}\else
  \'{n}\fi{}ski}\ \emph {et~al.}(2017)\citenamefont
  {Demkowicz-Dobrza\ifmmode~\acute{n}\else \'{n}\fi{}ski}, \citenamefont
  {Czajkowski},\ and\ \citenamefont {Sekatski}}]{PhysRevX.7.041009}%
  \BibitemOpen
  \bibfield  {author} {\bibinfo {author} {\bibfnamefont {R.}~\bibnamefont
  {Demkowicz-Dobrza\ifmmode~\acute{n}\else \'{n}\fi{}ski}}, \bibinfo {author}
  {\bibfnamefont {J.}~\bibnamefont {Czajkowski}}, \ and\ \bibinfo {author}
  {\bibfnamefont {P.}~\bibnamefont {Sekatski}},\ }\href {\doibase
  10.1103/PhysRevX.7.041009} {\bibfield  {journal} {\bibinfo  {journal} {Phys.
  Rev. X}\ }\textbf {\bibinfo {volume} {7}},\ \bibinfo {pages} {041009}
  (\bibinfo {year} {2017})}\BibitemShut {NoStop}%
\bibitem [{\citenamefont {Taylor}\ \emph {et~al.}(2008)\citenamefont {Taylor},
  \citenamefont {Cappellaro}, \citenamefont {Childress}, \citenamefont {Jiang},
  \citenamefont {Budker}, \citenamefont {Hemmer}, \citenamefont {Yacoby},
  \citenamefont {Walsworth},\ and\ \citenamefont {Lukin}}]{Taylor2008}%
  \BibitemOpen
  \bibfield  {author} {\bibinfo {author} {\bibfnamefont {J.~M.}\ \bibnamefont
  {Taylor}}, \bibinfo {author} {\bibfnamefont {P.}~\bibnamefont {Cappellaro}},
  \bibinfo {author} {\bibfnamefont {L.}~\bibnamefont {Childress}}, \bibinfo
  {author} {\bibfnamefont {L.}~\bibnamefont {Jiang}}, \bibinfo {author}
  {\bibfnamefont {D.}~\bibnamefont {Budker}}, \bibinfo {author} {\bibfnamefont
  {P.~R.}\ \bibnamefont {Hemmer}}, \bibinfo {author} {\bibfnamefont
  {A.}~\bibnamefont {Yacoby}}, \bibinfo {author} {\bibfnamefont
  {R.}~\bibnamefont {Walsworth}}, \ and\ \bibinfo {author} {\bibfnamefont
  {M.~D.}\ \bibnamefont {Lukin}},\ }\href {\doibase 10.1038/nphys1075}
  {\bibfield  {journal} {\bibinfo  {journal} {Nature Physics}\ }\textbf
  {\bibinfo {volume} {4}},\ \bibinfo {pages} {810} (\bibinfo {year}
  {2008})}\BibitemShut {NoStop}%
\bibitem [{\citenamefont {de~Lange}\ \emph {et~al.}(2011)\citenamefont
  {de~Lange}, \citenamefont {Rist\`e}, \citenamefont {Dobrovitski},\ and\
  \citenamefont {Hanson}}]{PhysRevLett.106.080802}%
  \BibitemOpen
  \bibfield  {author} {\bibinfo {author} {\bibfnamefont {G.}~\bibnamefont
  {de~Lange}}, \bibinfo {author} {\bibfnamefont {D.}~\bibnamefont {Rist\`e}},
  \bibinfo {author} {\bibfnamefont {V.~V.}\ \bibnamefont {Dobrovitski}}, \ and\
  \bibinfo {author} {\bibfnamefont {R.}~\bibnamefont {Hanson}},\ }\href
  {\doibase 10.1103/PhysRevLett.106.080802} {\bibfield  {journal} {\bibinfo
  {journal} {Phys. Rev. Lett.}\ }\textbf {\bibinfo {volume} {106}},\ \bibinfo
  {pages} {080802} (\bibinfo {year} {2011})}\BibitemShut {NoStop}%
\bibitem [{\citenamefont {Cohen}\ \emph {et~al.}(2016)\citenamefont {Cohen},
  \citenamefont {Pilnyak}, \citenamefont {Istrati}, \citenamefont {Retzker},\
  and\ \citenamefont {Eisenberg}}]{PhysRevA.94.012324}%
  \BibitemOpen
  \bibfield  {author} {\bibinfo {author} {\bibfnamefont {L.}~\bibnamefont
  {Cohen}}, \bibinfo {author} {\bibfnamefont {Y.}~\bibnamefont {Pilnyak}},
  \bibinfo {author} {\bibfnamefont {D.}~\bibnamefont {Istrati}}, \bibinfo
  {author} {\bibfnamefont {A.}~\bibnamefont {Retzker}}, \ and\ \bibinfo
  {author} {\bibfnamefont {H.~S.}\ \bibnamefont {Eisenberg}},\ }\href {\doibase
  10.1103/PhysRevA.94.012324} {\bibfield  {journal} {\bibinfo  {journal} {Phys.
  Rev. A}\ }\textbf {\bibinfo {volume} {94}},\ \bibinfo {pages} {012324}
  (\bibinfo {year} {2016})}\BibitemShut {NoStop}%
\bibitem [{\citenamefont {Unden}\ \emph {et~al.}(2016)\citenamefont {Unden},
  \citenamefont {Balasubramanian}, \citenamefont {Louzon}, \citenamefont
  {Vinkler}, \citenamefont {Plenio}, \citenamefont {Markham}, \citenamefont
  {Twitchen}, \citenamefont {Stacey}, \citenamefont {Lovchinsky}, \citenamefont
  {Sushkov}, \citenamefont {Lukin}, \citenamefont {Retzker}, \citenamefont
  {Naydenov}, \citenamefont {McGuinness},\ and\ \citenamefont
  {Jelezko}}]{PhysRevLett.116.230502}%
  \BibitemOpen
  \bibfield  {author} {\bibinfo {author} {\bibfnamefont {T.}~\bibnamefont
  {Unden}}, \bibinfo {author} {\bibfnamefont {P.}~\bibnamefont
  {Balasubramanian}}, \bibinfo {author} {\bibfnamefont {D.}~\bibnamefont
  {Louzon}}, \bibinfo {author} {\bibfnamefont {Y.}~\bibnamefont {Vinkler}},
  \bibinfo {author} {\bibfnamefont {M.~B.}\ \bibnamefont {Plenio}}, \bibinfo
  {author} {\bibfnamefont {M.}~\bibnamefont {Markham}}, \bibinfo {author}
  {\bibfnamefont {D.}~\bibnamefont {Twitchen}}, \bibinfo {author}
  {\bibfnamefont {A.}~\bibnamefont {Stacey}}, \bibinfo {author} {\bibfnamefont
  {I.}~\bibnamefont {Lovchinsky}}, \bibinfo {author} {\bibfnamefont {A.~O.}\
  \bibnamefont {Sushkov}}, \bibinfo {author} {\bibfnamefont {M.~D.}\
  \bibnamefont {Lukin}}, \bibinfo {author} {\bibfnamefont {A.}~\bibnamefont
  {Retzker}}, \bibinfo {author} {\bibfnamefont {B.}~\bibnamefont {Naydenov}},
  \bibinfo {author} {\bibfnamefont {L.~P.}\ \bibnamefont {McGuinness}}, \ and\
  \bibinfo {author} {\bibfnamefont {F.}~\bibnamefont {Jelezko}},\ }\href
  {\doibase 10.1103/PhysRevLett.116.230502} {\bibfield  {journal} {\bibinfo
  {journal} {Phys. Rev. Lett.}\ }\textbf {\bibinfo {volume} {116}},\ \bibinfo
  {pages} {230502} (\bibinfo {year} {2016})}\BibitemShut {NoStop}%
\bibitem [{\citenamefont {Ho}\ \emph {et~al.}(2019{\natexlab{a}})\citenamefont
  {Ho}, \citenamefont {Matsuzaki}, \citenamefont {Matsuzaki},\ and\
  \citenamefont {Kondo}}]{Ho_2019}%
  \BibitemOpen
  \bibfield  {author} {\bibinfo {author} {\bibfnamefont {L.~B.}\ \bibnamefont
  {Ho}}, \bibinfo {author} {\bibfnamefont {Y.}~\bibnamefont {Matsuzaki}},
  \bibinfo {author} {\bibfnamefont {M.}~\bibnamefont {Matsuzaki}}, \ and\
  \bibinfo {author} {\bibfnamefont {Y.}~\bibnamefont {Kondo}},\ }\href
  {\doibase 10.1088/1367-2630/ab3a25} {\bibfield  {journal} {\bibinfo
  {journal} {New Journal of Physics}\ }\textbf {\bibinfo {volume} {21}},\
  \bibinfo {pages} {093008} (\bibinfo {year} {2019}{\natexlab{a}})}\BibitemShut
  {NoStop}%
\bibitem [{\citenamefont {Ho}\ \emph {et~al.}(2019{\natexlab{b}})\citenamefont
  {Ho}, \citenamefont {Matsuzaki}, \citenamefont {Matsuzaki},\ and\
  \citenamefont {Kondo}}]{ho2019nuclear}%
  \BibitemOpen
  \bibfield  {author} {\bibinfo {author} {\bibfnamefont {L.~B.}\ \bibnamefont
  {Ho}}, \bibinfo {author} {\bibfnamefont {Y.}~\bibnamefont {Matsuzaki}},
  \bibinfo {author} {\bibfnamefont {M.}~\bibnamefont {Matsuzaki}}, \ and\
  \bibinfo {author} {\bibfnamefont {Y.}~\bibnamefont {Kondo}},\ }\href@noop {}
  {\enquote {\bibinfo {title} {Nuclear magnetic resonance model of an entangled
  sensor under noise},}\ } (\bibinfo {year} {2019}{\natexlab{b}}),\ \Eprint
  {http://arxiv.org/abs/1910.13599} {arXiv:1910.13599 [quant-ph]} \BibitemShut
  {NoStop}%
\bibitem [{\citenamefont {Chin}\ \emph {et~al.}(2012)\citenamefont {Chin},
  \citenamefont {Huelga},\ and\ \citenamefont {Plenio}}]{chin2012quantum}%
  \BibitemOpen
  \bibfield  {author} {\bibinfo {author} {\bibfnamefont {A.~W.}\ \bibnamefont
  {Chin}}, \bibinfo {author} {\bibfnamefont {S.~F.}\ \bibnamefont {Huelga}}, \
  and\ \bibinfo {author} {\bibfnamefont {M.~B.}\ \bibnamefont {Plenio}},\
  }\href@noop {} {\bibfield  {journal} {\bibinfo  {journal} {Physical review
  letters}\ }\textbf {\bibinfo {volume} {109}},\ \bibinfo {pages} {233601}
  (\bibinfo {year} {2012})}\BibitemShut {NoStop}%
\bibitem [{\citenamefont {Macieszczak}(2015)}]{macieszczak2015zeno}%
  \BibitemOpen
  \bibfield  {author} {\bibinfo {author} {\bibfnamefont {K.}~\bibnamefont
  {Macieszczak}},\ }\href@noop {} {\bibfield  {journal} {\bibinfo  {journal}
  {Physical Review A}\ }\textbf {\bibinfo {volume} {92}},\ \bibinfo {pages}
  {010102} (\bibinfo {year} {2015})}\BibitemShut {NoStop}%
\bibitem [{\citenamefont {Tanaka}\ \emph {et~al.}(2015)\citenamefont {Tanaka},
  \citenamefont {Knott}, \citenamefont {Matsuzaki}, \citenamefont {Dooley},
  \citenamefont {Yamaguchi}, \citenamefont {Munro},\ and\ \citenamefont
  {Saito}}]{tanaka2015proposed}%
  \BibitemOpen
  \bibfield  {author} {\bibinfo {author} {\bibfnamefont {T.}~\bibnamefont
  {Tanaka}}, \bibinfo {author} {\bibfnamefont {P.}~\bibnamefont {Knott}},
  \bibinfo {author} {\bibfnamefont {Y.}~\bibnamefont {Matsuzaki}}, \bibinfo
  {author} {\bibfnamefont {S.}~\bibnamefont {Dooley}}, \bibinfo {author}
  {\bibfnamefont {H.}~\bibnamefont {Yamaguchi}}, \bibinfo {author}
  {\bibfnamefont {W.~J.}\ \bibnamefont {Munro}}, \ and\ \bibinfo {author}
  {\bibfnamefont {S.}~\bibnamefont {Saito}},\ }\href@noop {} {\bibfield
  {journal} {\bibinfo  {journal} {Physical review letters}\ }\textbf {\bibinfo
  {volume} {115}},\ \bibinfo {pages} {170801} (\bibinfo {year}
  {2015})}\BibitemShut {NoStop}%
\bibitem [{\citenamefont {Smirne}\ \emph {et~al.}(2016)\citenamefont {Smirne},
  \citenamefont {Ko{\l}ody{\'n}ski}, \citenamefont {Huelga},\ and\
  \citenamefont {Demkowicz-Dobrza{\'n}ski}}]{smirne2016ultimate}%
  \BibitemOpen
  \bibfield  {author} {\bibinfo {author} {\bibfnamefont {A.}~\bibnamefont
  {Smirne}}, \bibinfo {author} {\bibfnamefont {J.}~\bibnamefont
  {Ko{\l}ody{\'n}ski}}, \bibinfo {author} {\bibfnamefont {S.~F.}\ \bibnamefont
  {Huelga}}, \ and\ \bibinfo {author} {\bibfnamefont {R.}~\bibnamefont
  {Demkowicz-Dobrza{\'n}ski}},\ }\href@noop {} {\bibfield  {journal} {\bibinfo
  {journal} {Physical review letters}\ }\textbf {\bibinfo {volume} {116}},\
  \bibinfo {pages} {120801} (\bibinfo {year} {2016})}\BibitemShut {NoStop}%
\bibitem [{\citenamefont {Haase}\ \emph {et~al.}(2018)\citenamefont {Haase},
  \citenamefont {Smirne}, \citenamefont {Ko{\l}ody{\'n}ski}, \citenamefont
  {Demkowicz-Dobrza{\'n}ski},\ and\ \citenamefont
  {Huelga}}]{haase2018fundamental}%
  \BibitemOpen
  \bibfield  {author} {\bibinfo {author} {\bibfnamefont {J.~F.}\ \bibnamefont
  {Haase}}, \bibinfo {author} {\bibfnamefont {A.}~\bibnamefont {Smirne}},
  \bibinfo {author} {\bibfnamefont {J.}~\bibnamefont {Ko{\l}ody{\'n}ski}},
  \bibinfo {author} {\bibfnamefont {R.}~\bibnamefont
  {Demkowicz-Dobrza{\'n}ski}}, \ and\ \bibinfo {author} {\bibfnamefont {S.~F.}\
  \bibnamefont {Huelga}},\ }\href@noop {} {\bibfield  {journal} {\bibinfo
  {journal} {New Journal of Physics}\ }\textbf {\bibinfo {volume} {20}},\
  \bibinfo {pages} {053009} (\bibinfo {year} {2018})}\BibitemShut {NoStop}%
\bibitem [{\citenamefont {Dooley}\ \emph {et~al.}(2016)\citenamefont {Dooley},
  \citenamefont {Yukawa}, \citenamefont {Matsuzaki}, \citenamefont {Knee},
  \citenamefont {Munro},\ and\ \citenamefont {Nemoto}}]{dooley2016hybrid}%
  \BibitemOpen
  \bibfield  {author} {\bibinfo {author} {\bibfnamefont {S.}~\bibnamefont
  {Dooley}}, \bibinfo {author} {\bibfnamefont {E.}~\bibnamefont {Yukawa}},
  \bibinfo {author} {\bibfnamefont {Y.}~\bibnamefont {Matsuzaki}}, \bibinfo
  {author} {\bibfnamefont {G.~C.}\ \bibnamefont {Knee}}, \bibinfo {author}
  {\bibfnamefont {W.~J.}\ \bibnamefont {Munro}}, \ and\ \bibinfo {author}
  {\bibfnamefont {K.}~\bibnamefont {Nemoto}},\ }\href@noop {} {\bibfield
  {journal} {\bibinfo  {journal} {New Journal of Physics}\ }\textbf {\bibinfo
  {volume} {18}},\ \bibinfo {pages} {053011} (\bibinfo {year}
  {2016})}\BibitemShut {NoStop}%
\bibitem [{\citenamefont {Albarelli}\ \emph {et~al.}(2019)\citenamefont
  {Albarelli}, \citenamefont {Friel},\ and\ \citenamefont
  {Datta}}]{PhysRevLett.123.200503}%
  \BibitemOpen
  \bibfield  {author} {\bibinfo {author} {\bibfnamefont {F.}~\bibnamefont
  {Albarelli}}, \bibinfo {author} {\bibfnamefont {J.~F.}\ \bibnamefont
  {Friel}}, \ and\ \bibinfo {author} {\bibfnamefont {A.}~\bibnamefont
  {Datta}},\ }\href {\doibase 10.1103/PhysRevLett.123.200503} {\bibfield
  {journal} {\bibinfo  {journal} {Phys. Rev. Lett.}\ }\textbf {\bibinfo
  {volume} {123}},\ \bibinfo {pages} {200503} (\bibinfo {year}
  {2019})}\BibitemShut {NoStop}%
\bibitem [{\citenamefont {Baumgratz}\ and\ \citenamefont
  {Datta}(2016)}]{PhysRevLett.116.030801}%
  \BibitemOpen
  \bibfield  {author} {\bibinfo {author} {\bibfnamefont {T.}~\bibnamefont
  {Baumgratz}}\ and\ \bibinfo {author} {\bibfnamefont {A.}~\bibnamefont
  {Datta}},\ }\href {\doibase 10.1103/PhysRevLett.116.030801} {\bibfield
  {journal} {\bibinfo  {journal} {Phys. Rev. Lett.}\ }\textbf {\bibinfo
  {volume} {116}},\ \bibinfo {pages} {030801} (\bibinfo {year}
  {2016})}\BibitemShut {NoStop}%
\bibitem [{\citenamefont {Vidrighin}\ \emph {et~al.}(2014)\citenamefont
  {Vidrighin}, \citenamefont {Donati}, \citenamefont {Genoni}, \citenamefont
  {Jin}, \citenamefont {Kolthammer}, \citenamefont {Kim}, \citenamefont
  {Datta}, \citenamefont {Barbieri},\ and\ \citenamefont
  {Walmsley}}]{Vidrighin2014}%
  \BibitemOpen
  \bibfield  {author} {\bibinfo {author} {\bibfnamefont {M.~D.}\ \bibnamefont
  {Vidrighin}}, \bibinfo {author} {\bibfnamefont {G.}~\bibnamefont {Donati}},
  \bibinfo {author} {\bibfnamefont {M.~G.}\ \bibnamefont {Genoni}}, \bibinfo
  {author} {\bibfnamefont {X.-M.}\ \bibnamefont {Jin}}, \bibinfo {author}
  {\bibfnamefont {W.~S.}\ \bibnamefont {Kolthammer}}, \bibinfo {author}
  {\bibfnamefont {M.~S.}\ \bibnamefont {Kim}}, \bibinfo {author} {\bibfnamefont
  {A.}~\bibnamefont {Datta}}, \bibinfo {author} {\bibfnamefont
  {M.}~\bibnamefont {Barbieri}}, \ and\ \bibinfo {author} {\bibfnamefont
  {I.~A.}\ \bibnamefont {Walmsley}},\ }\href {\doibase 10.1038/ncomms4532}
  {\bibfield  {journal} {\bibinfo  {journal} {Nature Communications}\ }\textbf
  {\bibinfo {volume} {5}},\ \bibinfo {pages} {3532} (\bibinfo {year}
  {2014})}\BibitemShut {NoStop}%
\bibitem [{\citenamefont {Altorio}\ \emph {et~al.}(2015)\citenamefont
  {Altorio}, \citenamefont {Genoni}, \citenamefont {Vidrighin}, \citenamefont
  {Somma},\ and\ \citenamefont {Barbieri}}]{PhysRevA.92.032114}%
  \BibitemOpen
  \bibfield  {author} {\bibinfo {author} {\bibfnamefont {M.}~\bibnamefont
  {Altorio}}, \bibinfo {author} {\bibfnamefont {M.~G.}\ \bibnamefont {Genoni}},
  \bibinfo {author} {\bibfnamefont {M.~D.}\ \bibnamefont {Vidrighin}}, \bibinfo
  {author} {\bibfnamefont {F.}~\bibnamefont {Somma}}, \ and\ \bibinfo {author}
  {\bibfnamefont {M.}~\bibnamefont {Barbieri}},\ }\href {\doibase
  10.1103/PhysRevA.92.032114} {\bibfield  {journal} {\bibinfo  {journal} {Phys.
  Rev. A}\ }\textbf {\bibinfo {volume} {92}},\ \bibinfo {pages} {032114}
  (\bibinfo {year} {2015})}\BibitemShut {NoStop}%
\bibitem [{\citenamefont {Szczykulska}\ \emph {et~al.}(2017)\citenamefont
  {Szczykulska}, \citenamefont {Baumgratz},\ and\ \citenamefont
  {Datta}}]{Szczykulska_2017}%
  \BibitemOpen
  \bibfield  {author} {\bibinfo {author} {\bibfnamefont {M.}~\bibnamefont
  {Szczykulska}}, \bibinfo {author} {\bibfnamefont {T.}~\bibnamefont
  {Baumgratz}}, \ and\ \bibinfo {author} {\bibfnamefont {A.}~\bibnamefont
  {Datta}},\ }\href {\doibase 10.1088/2058-9565/aa7fa9} {\bibfield  {journal}
  {\bibinfo  {journal} {Quantum Science and Technology}\ }\textbf {\bibinfo
  {volume} {2}},\ \bibinfo {pages} {044004} (\bibinfo {year}
  {2017})}\BibitemShut {NoStop}%
\bibitem [{\citenamefont {Knysh}\ and\ \citenamefont
  {Durkin}(2013)}]{knysh2013estimation}%
  \BibitemOpen
  \bibfield  {author} {\bibinfo {author} {\bibfnamefont {S.~I.}\ \bibnamefont
  {Knysh}}\ and\ \bibinfo {author} {\bibfnamefont {G.~A.}\ \bibnamefont
  {Durkin}},\ }\href@noop {} {\enquote {\bibinfo {title} {Estimation of phase
  and diffusion: Combining quantum statistics and classical noise},}\ }
  (\bibinfo {year} {2013}),\ \Eprint {http://arxiv.org/abs/1307.0470}
  {arXiv:1307.0470 [quant-ph]} \BibitemShut {NoStop}%
\bibitem [{\citenamefont {Crowley}\ \emph {et~al.}(2014)\citenamefont
  {Crowley}, \citenamefont {Datta}, \citenamefont {Barbieri},\ and\
  \citenamefont {Walmsley}}]{PhysRevA.89.023845}%
  \BibitemOpen
  \bibfield  {author} {\bibinfo {author} {\bibfnamefont {P.~J.~D.}\
  \bibnamefont {Crowley}}, \bibinfo {author} {\bibfnamefont {A.}~\bibnamefont
  {Datta}}, \bibinfo {author} {\bibfnamefont {M.}~\bibnamefont {Barbieri}}, \
  and\ \bibinfo {author} {\bibfnamefont {I.~A.}\ \bibnamefont {Walmsley}},\
  }\href {\doibase 10.1103/PhysRevA.89.023845} {\bibfield  {journal} {\bibinfo
  {journal} {Phys. Rev. A}\ }\textbf {\bibinfo {volume} {89}},\ \bibinfo
  {pages} {023845} (\bibinfo {year} {2014})}\BibitemShut {NoStop}%
\bibitem [{\citenamefont {Pinel}\ \emph {et~al.}(2013)\citenamefont {Pinel},
  \citenamefont {Jian}, \citenamefont {Treps}, \citenamefont {Fabre},\ and\
  \citenamefont {Braun}}]{PhysRevA.88.040102}%
  \BibitemOpen
  \bibfield  {author} {\bibinfo {author} {\bibfnamefont {O.}~\bibnamefont
  {Pinel}}, \bibinfo {author} {\bibfnamefont {P.}~\bibnamefont {Jian}},
  \bibinfo {author} {\bibfnamefont {N.}~\bibnamefont {Treps}}, \bibinfo
  {author} {\bibfnamefont {C.}~\bibnamefont {Fabre}}, \ and\ \bibinfo {author}
  {\bibfnamefont {D.}~\bibnamefont {Braun}},\ }\href {\doibase
  10.1103/PhysRevA.88.040102} {\bibfield  {journal} {\bibinfo  {journal} {Phys.
  Rev. A}\ }\textbf {\bibinfo {volume} {88}},\ \bibinfo {pages} {040102}
  (\bibinfo {year} {2013})}\BibitemShut {NoStop}%
\bibitem [{\citenamefont {Gagatsos}\ \emph {et~al.}(2017)\citenamefont
  {Gagatsos}, \citenamefont {Bash}, \citenamefont {Guha},\ and\ \citenamefont
  {Datta}}]{PhysRevA.96.062306}%
  \BibitemOpen
  \bibfield  {author} {\bibinfo {author} {\bibfnamefont {C.~N.}\ \bibnamefont
  {Gagatsos}}, \bibinfo {author} {\bibfnamefont {B.~A.}\ \bibnamefont {Bash}},
  \bibinfo {author} {\bibfnamefont {S.}~\bibnamefont {Guha}}, \ and\ \bibinfo
  {author} {\bibfnamefont {A.}~\bibnamefont {Datta}},\ }\href {\doibase
  10.1103/PhysRevA.96.062306} {\bibfield  {journal} {\bibinfo  {journal} {Phys.
  Rev. A}\ }\textbf {\bibinfo {volume} {96}},\ \bibinfo {pages} {062306}
  (\bibinfo {year} {2017})}\BibitemShut {NoStop}%
\bibitem [{\citenamefont {Roccia}\ \emph {et~al.}(2018)\citenamefont {Roccia},
  \citenamefont {Cimini}, \citenamefont {Sbroscia}, \citenamefont {Gianani},
  \citenamefont {Ruggiero}, \citenamefont {Mancino}, \citenamefont {Genoni},
  \citenamefont {Ricci},\ and\ \citenamefont {Barbieri}}]{Roccia:18}%
  \BibitemOpen
  \bibfield  {author} {\bibinfo {author} {\bibfnamefont {E.}~\bibnamefont
  {Roccia}}, \bibinfo {author} {\bibfnamefont {V.}~\bibnamefont {Cimini}},
  \bibinfo {author} {\bibfnamefont {M.}~\bibnamefont {Sbroscia}}, \bibinfo
  {author} {\bibfnamefont {I.}~\bibnamefont {Gianani}}, \bibinfo {author}
  {\bibfnamefont {L.}~\bibnamefont {Ruggiero}}, \bibinfo {author}
  {\bibfnamefont {L.}~\bibnamefont {Mancino}}, \bibinfo {author} {\bibfnamefont
  {M.~G.}\ \bibnamefont {Genoni}}, \bibinfo {author} {\bibfnamefont {M.~A.}\
  \bibnamefont {Ricci}}, \ and\ \bibinfo {author} {\bibfnamefont
  {M.}~\bibnamefont {Barbieri}},\ }\href {\doibase 10.1364/OPTICA.5.001171}
  {\bibfield  {journal} {\bibinfo  {journal} {Optica}\ }\textbf {\bibinfo
  {volume} {5}},\ \bibinfo {pages} {1171} (\bibinfo {year} {2018})}\BibitemShut
  {NoStop}%
\bibitem [{\citenamefont {Genoni}\ \emph {et~al.}(2013)\citenamefont {Genoni},
  \citenamefont {Paris}, \citenamefont {Adesso}, \citenamefont {Nha},
  \citenamefont {Knight},\ and\ \citenamefont {Kim}}]{PhysRevA.87.012107}%
  \BibitemOpen
  \bibfield  {author} {\bibinfo {author} {\bibfnamefont {M.~G.}\ \bibnamefont
  {Genoni}}, \bibinfo {author} {\bibfnamefont {M.~G.~A.}\ \bibnamefont
  {Paris}}, \bibinfo {author} {\bibfnamefont {G.}~\bibnamefont {Adesso}},
  \bibinfo {author} {\bibfnamefont {H.}~\bibnamefont {Nha}}, \bibinfo {author}
  {\bibfnamefont {P.~L.}\ \bibnamefont {Knight}}, \ and\ \bibinfo {author}
  {\bibfnamefont {M.~S.}\ \bibnamefont {Kim}},\ }\href {\doibase
  10.1103/PhysRevA.87.012107} {\bibfield  {journal} {\bibinfo  {journal} {Phys.
  Rev. A}\ }\textbf {\bibinfo {volume} {87}},\ \bibinfo {pages} {012107}
  (\bibinfo {year} {2013})}\BibitemShut {NoStop}%
\bibitem [{\citenamefont {Steinlechner}\ \emph {et~al.}(2013)\citenamefont
  {Steinlechner}, \citenamefont {Bauchrowitz}, \citenamefont {Meinders},
  \citenamefont {M{\"u}ller-Ebhardt}, \citenamefont {Danzmann},\ and\
  \citenamefont {Schnabel}}]{Steinlechner2013}%
  \BibitemOpen
  \bibfield  {author} {\bibinfo {author} {\bibfnamefont {S.}~\bibnamefont
  {Steinlechner}}, \bibinfo {author} {\bibfnamefont {J.}~\bibnamefont
  {Bauchrowitz}}, \bibinfo {author} {\bibfnamefont {M.}~\bibnamefont
  {Meinders}}, \bibinfo {author} {\bibfnamefont {H.}~\bibnamefont
  {M{\"u}ller-Ebhardt}}, \bibinfo {author} {\bibfnamefont {K.}~\bibnamefont
  {Danzmann}}, \ and\ \bibinfo {author} {\bibfnamefont {R.}~\bibnamefont
  {Schnabel}},\ }\href {\doibase 10.1038/nphoton.2013.150} {\bibfield
  {journal} {\bibinfo  {journal} {Nature Photonics}\ }\textbf {\bibinfo
  {volume} {7}},\ \bibinfo {pages} {626} (\bibinfo {year} {2013})}\BibitemShut
  {NoStop}%
\bibitem [{\citenamefont {Cyril}\ \emph {et~al.}(2013)\citenamefont {Cyril},
  \citenamefont {Tommaso},\ and\ \citenamefont {G.}}]{Cyril2013}%
  \BibitemOpen
  \bibfield  {author} {\bibinfo {author} {\bibfnamefont {V.}~\bibnamefont
  {Cyril}}, \bibinfo {author} {\bibfnamefont {T.}~\bibnamefont {Tommaso}}, \
  and\ \bibinfo {author} {\bibfnamefont {G.~M.}\ \bibnamefont {G.}},\ }\enquote
  {\bibinfo {title} {qmetro},}\ \ (\bibinfo {year} {2013})\ Chap.\ \bibinfo
  {chapter} {Quantum estimation of a two-phase spin rotation}, p.~\bibinfo
  {pages} {12}\BibitemShut {NoStop}%
\bibitem [{\citenamefont {Humphreys}\ \emph {et~al.}(2013)\citenamefont
  {Humphreys}, \citenamefont {Barbieri}, \citenamefont {Datta},\ and\
  \citenamefont {Walmsley}}]{PhysRevLett.111.070403}%
  \BibitemOpen
  \bibfield  {author} {\bibinfo {author} {\bibfnamefont {P.~C.}\ \bibnamefont
  {Humphreys}}, \bibinfo {author} {\bibfnamefont {M.}~\bibnamefont {Barbieri}},
  \bibinfo {author} {\bibfnamefont {A.}~\bibnamefont {Datta}}, \ and\ \bibinfo
  {author} {\bibfnamefont {I.~A.}\ \bibnamefont {Walmsley}},\ }\href {\doibase
  10.1103/PhysRevLett.111.070403} {\bibfield  {journal} {\bibinfo  {journal}
  {Phys. Rev. Lett.}\ }\textbf {\bibinfo {volume} {111}},\ \bibinfo {pages}
  {070403} (\bibinfo {year} {2013})}\BibitemShut {NoStop}%
\bibitem [{\citenamefont {Liu}\ and\ \citenamefont {Cable}(2017)}]{Liu_2017}%
  \BibitemOpen
  \bibfield  {author} {\bibinfo {author} {\bibfnamefont {N.}~\bibnamefont
  {Liu}}\ and\ \bibinfo {author} {\bibfnamefont {H.}~\bibnamefont {Cable}},\
  }\href {\doibase 10.1088/2058-9565/aa6fea} {\bibfield  {journal} {\bibinfo
  {journal} {Quantum Science and Technology}\ }\textbf {\bibinfo {volume}
  {2}},\ \bibinfo {pages} {025008} (\bibinfo {year} {2017})}\BibitemShut
  {NoStop}%
\bibitem [{\citenamefont {Monras}\ and\ \citenamefont
  {Illuminati}(2011)}]{PhysRevA.83.012315}%
  \BibitemOpen
  \bibfield  {author} {\bibinfo {author} {\bibfnamefont {A.}~\bibnamefont
  {Monras}}\ and\ \bibinfo {author} {\bibfnamefont {F.}~\bibnamefont
  {Illuminati}},\ }\href {\doibase 10.1103/PhysRevA.83.012315} {\bibfield
  {journal} {\bibinfo  {journal} {Phys. Rev. A}\ }\textbf {\bibinfo {volume}
  {83}},\ \bibinfo {pages} {012315} (\bibinfo {year} {2011})}\BibitemShut
  {NoStop}%
\bibitem [{\citenamefont {Berry}\ \emph {et~al.}(2015)\citenamefont {Berry},
  \citenamefont {Tsang}, \citenamefont {Hall},\ and\ \citenamefont
  {Wiseman}}]{PhysRevX.5.031018}%
  \BibitemOpen
  \bibfield  {author} {\bibinfo {author} {\bibfnamefont {D.~W.}\ \bibnamefont
  {Berry}}, \bibinfo {author} {\bibfnamefont {M.}~\bibnamefont {Tsang}},
  \bibinfo {author} {\bibfnamefont {M.~J.~W.}\ \bibnamefont {Hall}}, \ and\
  \bibinfo {author} {\bibfnamefont {H.~M.}\ \bibnamefont {Wiseman}},\ }\href
  {\doibase 10.1103/PhysRevX.5.031018} {\bibfield  {journal} {\bibinfo
  {journal} {Phys. Rev. X}\ }\textbf {\bibinfo {volume} {5}},\ \bibinfo {pages}
  {031018} (\bibinfo {year} {2015})}\BibitemShut {NoStop}%
\bibitem [{\citenamefont {Fujiwara}(2001)}]{PhysRevA.65.012316}%
  \BibitemOpen
  \bibfield  {author} {\bibinfo {author} {\bibfnamefont {A.}~\bibnamefont
  {Fujiwara}},\ }\href {\doibase 10.1103/PhysRevA.65.012316} {\bibfield
  {journal} {\bibinfo  {journal} {Phys. Rev. A}\ }\textbf {\bibinfo {volume}
  {65}},\ \bibinfo {pages} {012316} (\bibinfo {year} {2001})}\BibitemShut
  {NoStop}%
\bibitem [{\citenamefont {Ballester}(2004)}]{PhysRevA.69.022303}%
  \BibitemOpen
  \bibfield  {author} {\bibinfo {author} {\bibfnamefont {M.~A.}\ \bibnamefont
  {Ballester}},\ }\href {\doibase 10.1103/PhysRevA.69.022303} {\bibfield
  {journal} {\bibinfo  {journal} {Phys. Rev. A}\ }\textbf {\bibinfo {volume}
  {69}},\ \bibinfo {pages} {022303} (\bibinfo {year} {2004})}\BibitemShut
  {NoStop}%
\bibitem [{\citenamefont {Le~Sage}\ \emph {et~al.}(2013)\citenamefont
  {Le~Sage}, \citenamefont {Arai}, \citenamefont {Glenn}, \citenamefont
  {DeVience}, \citenamefont {Pham}, \citenamefont {Rahn-Lee}, \citenamefont
  {Lukin}, \citenamefont {Yacoby}, \citenamefont {Komeili},\ and\ \citenamefont
  {Walsworth}}]{le2013optical}%
  \BibitemOpen
  \bibfield  {author} {\bibinfo {author} {\bibfnamefont {D.}~\bibnamefont
  {Le~Sage}}, \bibinfo {author} {\bibfnamefont {K.}~\bibnamefont {Arai}},
  \bibinfo {author} {\bibfnamefont {D.~R.}\ \bibnamefont {Glenn}}, \bibinfo
  {author} {\bibfnamefont {S.~J.}\ \bibnamefont {DeVience}}, \bibinfo {author}
  {\bibfnamefont {L.~M.}\ \bibnamefont {Pham}}, \bibinfo {author}
  {\bibfnamefont {L.}~\bibnamefont {Rahn-Lee}}, \bibinfo {author}
  {\bibfnamefont {M.~D.}\ \bibnamefont {Lukin}}, \bibinfo {author}
  {\bibfnamefont {A.}~\bibnamefont {Yacoby}}, \bibinfo {author} {\bibfnamefont
  {A.}~\bibnamefont {Komeili}}, \ and\ \bibinfo {author} {\bibfnamefont
  {R.~L.}\ \bibnamefont {Walsworth}},\ }\href@noop {} {\bibfield  {journal}
  {\bibinfo  {journal} {Nature}\ }\textbf {\bibinfo {volume} {496}},\ \bibinfo
  {pages} {486} (\bibinfo {year} {2013})}\BibitemShut {NoStop}%
\bibitem [{\citenamefont {Nowodzinski}\ \emph {et~al.}(2015)\citenamefont
  {Nowodzinski}, \citenamefont {Chipaux}, \citenamefont {Toraille},
  \citenamefont {Jacques}, \citenamefont {Roch},\ and\ \citenamefont
  {Debuisschert}}]{nowodzinski2015nitrogen}%
  \BibitemOpen
  \bibfield  {author} {\bibinfo {author} {\bibfnamefont {A.}~\bibnamefont
  {Nowodzinski}}, \bibinfo {author} {\bibfnamefont {M.}~\bibnamefont
  {Chipaux}}, \bibinfo {author} {\bibfnamefont {L.}~\bibnamefont {Toraille}},
  \bibinfo {author} {\bibfnamefont {V.}~\bibnamefont {Jacques}}, \bibinfo
  {author} {\bibfnamefont {J.-F.}\ \bibnamefont {Roch}}, \ and\ \bibinfo
  {author} {\bibfnamefont {T.}~\bibnamefont {Debuisschert}},\ }\href@noop {}
  {\bibfield  {journal} {\bibinfo  {journal} {Microelectronics Reliability}\
  }\textbf {\bibinfo {volume} {55}},\ \bibinfo {pages} {1549} (\bibinfo {year}
  {2015})}\BibitemShut {NoStop}%
\bibitem [{\citenamefont {Shankar}\ \emph {et~al.}(2017)\citenamefont
  {Shankar}, \citenamefont {Cooper}, \citenamefont {Bohnet}, \citenamefont
  {Bollinger},\ and\ \citenamefont {Holland}}]{shankar2017steady}%
  \BibitemOpen
  \bibfield  {author} {\bibinfo {author} {\bibfnamefont {A.}~\bibnamefont
  {Shankar}}, \bibinfo {author} {\bibfnamefont {J.}~\bibnamefont {Cooper}},
  \bibinfo {author} {\bibfnamefont {J.~G.}\ \bibnamefont {Bohnet}}, \bibinfo
  {author} {\bibfnamefont {J.~J.}\ \bibnamefont {Bollinger}}, \ and\ \bibinfo
  {author} {\bibfnamefont {M.}~\bibnamefont {Holland}},\ }\href@noop {}
  {\bibfield  {journal} {\bibinfo  {journal} {Physical Review A}\ }\textbf
  {\bibinfo {volume} {95}},\ \bibinfo {pages} {033423} (\bibinfo {year}
  {2017})}\BibitemShut {NoStop}%
\bibitem [{\citenamefont {Kirton}\ and\ \citenamefont
  {Keeling}(2017)}]{kirton2017suppressing}%
  \BibitemOpen
  \bibfield  {author} {\bibinfo {author} {\bibfnamefont {P.}~\bibnamefont
  {Kirton}}\ and\ \bibinfo {author} {\bibfnamefont {J.}~\bibnamefont
  {Keeling}},\ }\href@noop {} {\bibfield  {journal} {\bibinfo  {journal}
  {Physical review letters}\ }\textbf {\bibinfo {volume} {118}},\ \bibinfo
  {pages} {123602} (\bibinfo {year} {2017})}\BibitemShut {NoStop}%
\bibitem [{\citenamefont {Shammah}\ \emph {et~al.}(2018)\citenamefont
  {Shammah}, \citenamefont {Ahmed}, \citenamefont {Lambert}, \citenamefont
  {De~Liberato},\ and\ \citenamefont {Nori}}]{PhysRevA.98.063815}%
  \BibitemOpen
  \bibfield  {author} {\bibinfo {author} {\bibfnamefont {N.}~\bibnamefont
  {Shammah}}, \bibinfo {author} {\bibfnamefont {S.}~\bibnamefont {Ahmed}},
  \bibinfo {author} {\bibfnamefont {N.}~\bibnamefont {Lambert}}, \bibinfo
  {author} {\bibfnamefont {S.}~\bibnamefont {De~Liberato}}, \ and\ \bibinfo
  {author} {\bibfnamefont {F.}~\bibnamefont {Nori}},\ }\href {\doibase
  10.1103/PhysRevA.98.063815} {\bibfield  {journal} {\bibinfo  {journal} {Phys.
  Rev. A}\ }\textbf {\bibinfo {volume} {98}},\ \bibinfo {pages} {063815}
  (\bibinfo {year} {2018})}\BibitemShut {NoStop}%
\bibitem [{\citenamefont {Chru{\'s}ci{\'n}ski}\ and\ \citenamefont
  {Pascazio}(2017)}]{doi:10.1142/S1230161217400017}%
  \BibitemOpen
  \bibfield  {author} {\bibinfo {author} {\bibfnamefont {D.}~\bibnamefont
  {Chru{\'s}ci{\'n}ski}}\ and\ \bibinfo {author} {\bibfnamefont
  {S.}~\bibnamefont {Pascazio}},\ }\href {\doibase 10.1142/S1230161217400017}
  {\bibfield  {journal} {\bibinfo  {journal} {Open Systems \& Information
  Dynamics}\ }\textbf {\bibinfo {volume} {24}},\ \bibinfo {pages} {1740001}
  (\bibinfo {year} {2017})},\ \Eprint
  {http://arxiv.org/abs/https://doi.org/10.1142/S1230161217400017}
  {https://doi.org/10.1142/S1230161217400017} \BibitemShut {NoStop}%
\bibitem [{\citenamefont {HUSIMI}(1940)}]{1940264}%
  \BibitemOpen
  \bibfield  {author} {\bibinfo {author} {\bibfnamefont {K.}~\bibnamefont
  {HUSIMI}},\ }\href {\doibase 10.11429/ppmsj1919.22.4_264} {\bibfield
  {journal} {\bibinfo  {journal} {Proceedings of the Physico-Mathematical
  Society of Japan. 3rd Series}\ }\textbf {\bibinfo {volume} {22}},\ \bibinfo
  {pages} {264} (\bibinfo {year} {1940})}\BibitemShut {NoStop}%
\bibitem [{\citenamefont {Gorecki}\ \emph {et~al.}(2019)\citenamefont
  {Gorecki}, \citenamefont {Zhou}, \citenamefont {Jiang},\ and\ \citenamefont
  {Demkowicz-Dobrzanski}}]{gorecki2019optimal}%
  \BibitemOpen
  \bibfield  {author} {\bibinfo {author} {\bibfnamefont {W.}~\bibnamefont
  {Gorecki}}, \bibinfo {author} {\bibfnamefont {S.}~\bibnamefont {Zhou}},
  \bibinfo {author} {\bibfnamefont {L.}~\bibnamefont {Jiang}}, \ and\ \bibinfo
  {author} {\bibfnamefont {R.}~\bibnamefont {Demkowicz-Dobrzanski}},\
  }\href@noop {} {\enquote {\bibinfo {title} {Optimal probes and
  error-correction schemes in multi-parameter quantum metrology},}\ } (\bibinfo
  {year} {2019}),\ \Eprint {http://arxiv.org/abs/1901.00896} {arXiv:1901.00896
  [quant-ph]} \BibitemShut {NoStop}%
\bibitem [{\citenamefont {Chase}\ and\ \citenamefont
  {Geremia}(2008)}]{PhysRevA.78.052101}%
  \BibitemOpen
  \bibfield  {author} {\bibinfo {author} {\bibfnamefont {B.~A.}\ \bibnamefont
  {Chase}}\ and\ \bibinfo {author} {\bibfnamefont {J.~M.}\ \bibnamefont
  {Geremia}},\ }\href {\doibase 10.1103/PhysRevA.78.052101} {\bibfield
  {journal} {\bibinfo  {journal} {Phys. Rev. A}\ }\textbf {\bibinfo {volume}
  {78}},\ \bibinfo {pages} {052101} (\bibinfo {year} {2008})}\BibitemShut
  {NoStop}%
\bibitem [{\citenamefont {Baragiola}\ \emph {et~al.}(2010)\citenamefont
  {Baragiola}, \citenamefont {Chase},\ and\ \citenamefont
  {Geremia}}]{PhysRevA.81.032104}%
  \BibitemOpen
  \bibfield  {author} {\bibinfo {author} {\bibfnamefont {B.~Q.}\ \bibnamefont
  {Baragiola}}, \bibinfo {author} {\bibfnamefont {B.~A.}\ \bibnamefont
  {Chase}}, \ and\ \bibinfo {author} {\bibfnamefont {J.}~\bibnamefont
  {Geremia}},\ }\href {\doibase 10.1103/PhysRevA.81.032104} {\bibfield
  {journal} {\bibinfo  {journal} {Phys. Rev. A}\ }\textbf {\bibinfo {volume}
  {81}},\ \bibinfo {pages} {032104} (\bibinfo {year} {2010})}\BibitemShut
  {NoStop}%
\bibitem [{\citenamefont {Mihailov}(1977)}]{Mihailov_1977}%
  \BibitemOpen
  \bibfield  {author} {\bibinfo {author} {\bibfnamefont {V.~V.}\ \bibnamefont
  {Mihailov}},\ }\href {\doibase 10.1088/0305-4470/10/2/003} {\bibfield
  {journal} {\bibinfo  {journal} {Journal of Physics A: Mathematical and
  General}\ }\textbf {\bibinfo {volume} {10}},\ \bibinfo {pages} {147}
  (\bibinfo {year} {1977})}\BibitemShut {NoStop}%
\bibitem [{\citenamefont {Dicke}(1954)}]{PhysRev.93.99}%
  \BibitemOpen
  \bibfield  {author} {\bibinfo {author} {\bibfnamefont {R.~H.}\ \bibnamefont
  {Dicke}},\ }\href {\doibase 10.1103/PhysRev.93.99} {\bibfield  {journal}
  {\bibinfo  {journal} {Phys. Rev.}\ }\textbf {\bibinfo {volume} {93}},\
  \bibinfo {pages} {99} (\bibinfo {year} {1954})}\BibitemShut {NoStop}%
\bibitem [{\citenamefont {Wilcox}(1967)}]{doi:10.1063/1.1705306}%
  \BibitemOpen
  \bibfield  {author} {\bibinfo {author} {\bibfnamefont {R.~M.}\ \bibnamefont
  {Wilcox}},\ }\href {\doibase 10.1063/1.1705306} {\bibfield  {journal}
  {\bibinfo  {journal} {Journal of Mathematical Physics}\ }\textbf {\bibinfo
  {volume} {8}},\ \bibinfo {pages} {962} (\bibinfo {year} {1967})},\ \Eprint
  {http://arxiv.org/abs/https://doi.org/10.1063/1.1705306}
  {https://doi.org/10.1063/1.1705306} \BibitemShut {NoStop}%
\bibitem [{\citenamefont {Pang}\ and\ \citenamefont
  {Brun}(2014)}]{PhysRevA.90.022117}%
  \BibitemOpen
  \bibfield  {author} {\bibinfo {author} {\bibfnamefont {S.}~\bibnamefont
  {Pang}}\ and\ \bibinfo {author} {\bibfnamefont {T.~A.}\ \bibnamefont
  {Brun}},\ }\href {\doibase 10.1103/PhysRevA.90.022117} {\bibfield  {journal}
  {\bibinfo  {journal} {Phys. Rev. A}\ }\textbf {\bibinfo {volume} {90}},\
  \bibinfo {pages} {022117} (\bibinfo {year} {2014})}\BibitemShut {NoStop}%
\end{thebibliography}%
\end{document}